\documentstyle[sprocl,epsfig]{article}
\def\beq{\begin{equation}}
\def\eeq{\end{equation}}
\def\beqn{\begin{eqnarray}}
\def\eeqn{\end{eqnarray}}
\begin{document}
\title{THE STANDARD MODEL \\
INTERMEDIATE MASS HIGGS BOSON}
\author{ S. DAWSON}
\address{Physics Department, Brookhaven National Laboratory,\\
Upton, NY 11973, USA}
\maketitle\abstracts{
We consider the phenomenology of the Standard Model intermediate
mass Higgs boson, $71~GeV < M_h < 2 M_W$.  The motivation for a
Higgs boson in this mass region is emphasized.  The branching ratios
for the Higgs boson, including electroweak and QCD radiative 
corrections,
are presented, along with production cross sections for 
 $e^+e^-, \mu^+\mu^-, \gamma\gamma$, 
and hadronic interactions.  Search strategies are surveyed 
briefly.
}
\section{Introduction}
  
The search for the Higgs boson is one of the most important objectives  
of present and future  colliders.
A Higgs boson or some object like it is needed in order to give the
 $W^\pm$ and $Z$
gauge bosons their observed masses and to cancel divergences which arise
when radiative corrections to electroweak observables are computed.
Beyond the mere fact of its existence, however, we have few clues
as to the expected mass of the Higgs boson, which ${\it a~priori}$ is a
free parameter of the theory.  It is hence vital to be able to search through
all mass regimes.

Direct experimental searches for the Higgs boson at LEP and LEPII 
yield  the limit,~\cite{leplim,lep2lim}  
\beq  
M_h>70.7~GeV,
\eeq
with no room for a very light Higgs boson.  
On the other end of the scale,
there are theoretical indications that the simplest version of
the Standard Model is inconsistent if the Higgs boson is too
heavy.  Lattice calculations~\cite{lat}
 suggest $M_h < 700~GeV$,
while vector boson scattering~\cite{unit}
 violates unitarity unless $M_h < 
800~GeV$.  
These are not rigid upper limits, but instead imply that if a
Higgs boson is not found below around $700-800~GeV$ then
the Standard Model
 of electroweak interactions must be enlarged to
include some more complicated theory.

In this chapter, we discuss the physics of the intermediate
mass Higgs boson.  For our purposes, we will define this to
be a Higgs boson  which is heavier
than the current experimental limit, but 
which is too light to decay to a $W^+W^-$ pair and
so we consider,
\beq
71~GeV< M_h < 160~GeV
\quad . 
\eeq
A Higgs boson in this mass range decays almost entirely to $ b
{\overline b}$ pairs and  much of the phenomenology
of the intermediate mass Higgs search is 
focused on identifying $b$ quarks.
Once the $2W$ threshold is reached, the search strategies for
the Higgs boson are completely different from the 
intermediate mass case, since the dominant decay mode becomes
the decay to vector boson pairs.  
A weakly interacting
Higgs boson with $M_h>2 M_W$ can almost certainly be discovered
at the LHC, as is discussed in the chapter by J. Gunion.
It turns out that the intermediate mass region is the
most challenging to probe experimentally.  
  
There is considerable theoretical motivation for a Higgs boson
in the $100-200~GeV$ mass region.  A  compelling argument is
that in the minimal  version
 of the supersymmetric Standard Model (MSSM), the
lightest Higgs boson~\cite{susymass}
 must be lighter than around $130~GeV$.
In fact, any SUSY model which contains only  $SU(2)_L$
scalar doublets
and singlets and remains  perturbative up to the Planck scale must
have a neutral Higgs boson lighter than around $150~GeV$.\cite{kkw}
  So although
we consider only the Standard Model Higgs boson in this chapter, the
motivation for an intermediate mass Higgs boson  is considerably broader  
than the Standard Model alone.    If a Higgs boson in the intermediate
mass range were to be discovered, the great experimental challenge
would be to determine if it were a Standard Model Higgs boson or
one of the
neutral Higgs bosons associated with a SUSY model or some other 
extension of the Standard Model.

The consistency of the Standard Model 
gives us information  about the allowed
mass range for the Higgs boson.  The scalar potential
for a complex
 $SU(2)_L$ Higgs doublet $\phi$ is given by,\footnote{
A complete discussion of limits on the Higgs boson mass from the scalar 
potential is found in the chapter by M. Quiros.}
\beq
V=-\mu^2\mid \phi\mid^2+\lambda(\mid \phi\mid^2)^2
\quad .
\eeq
After spontaneous symmetry breaking, the 
neutral component of the Higgs doublet
acquires a vacuum expectation value (vev),
$\langle \phi^0\rangle =v/\sqrt{2}$, where
the scale of the
vev is set by muon decay to be $v=(\sqrt{2}G_F)^{-1/2}=
246~GeV$.   The quartic
coupling can then be related to the 
physical Higgs boson mass,
\beq
\lambda={M_h^2\over 2 v^2}
\quad .
\eeq
 
As with all couplings in a gauge theory, the quartic coupling, $\lambda$,
evolves with the energy scale, $Q$.  For large $\lambda$,
(and neglecting the effects of gauge 
couplings and the top quark), the evolution is roughly,
\beq
Q{d\lambda\over d Q}\sim {3\lambda^2\over 4 \pi^2}
\quad .
\eeq
This equation is easily solved for $\lambda(Q)$,
\beq
{1\over \lambda(Q)}\sim {1\over \lambda(\Lambda)}-
{3\over 4 \pi^2}\log\biggl({Q\over \Lambda}\biggr).  
\eeq
 The requirement that the
coupling, $\lambda$,
 be finite up to some large scale, $\Lambda$, 
(${\it i.e.}$ that there be no Landau pole), then
gives an upper bound on the Higgs mass,\footnote{
The requirement that the theory remain weakly interacting
below the scale $\Lambda$, $\lambda(\Lambda)/(4\pi)<1$, can
also be used to obtain an upper limit on the Higgs boson mass.  
In practice, the scale at which the theory becomes strongly
interacting is not so different from the scale at which
the theory develops a Landau pole.}
\beq
M_h^2<{8 \pi^2 v^2\over 3 \log(\Lambda/M_h)}  
\sim{8 \pi^2 v^2\over 3 \log(\Lambda/v)}  
\quad .
\eeq 
This bound is the solid line in Fig. 1.
We see that if $\Lambda$ goes to infinity, the Higgs boson
mass goes to zero, along with $\lambda$, and so
the theory becomes non-interacting or trivial.  
More sophisticated analyses~\cite{triv}
 include the running of all  gauge and Yukawa couplings,
but yield a similar upper bound on the Higgs mass as a function
of  $\Lambda$. The scale $\Lambda$ can be thought of as the
scale at which new physics arises and the Standard Model is
no longer a  valid theory.  

\begin{figure}[tb]
\centerline{\epsfig{file=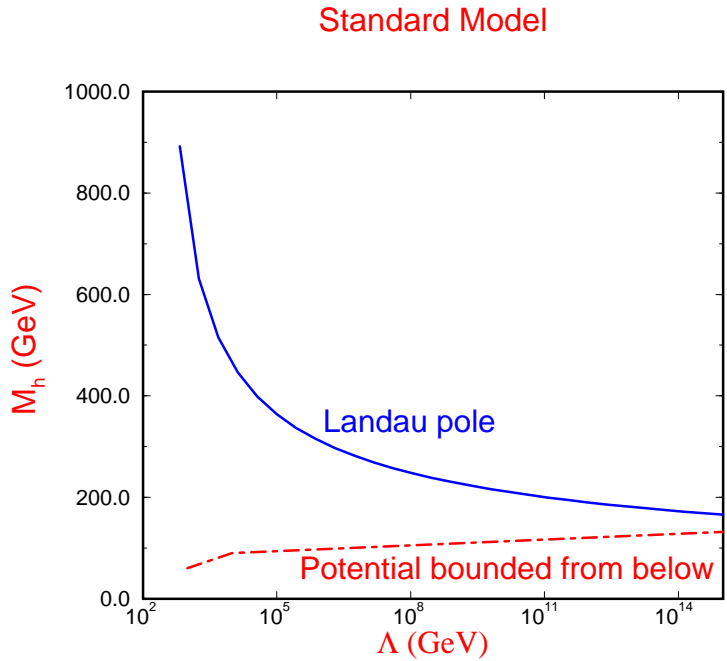,height=3.5in}}
\caption{ Theoretically allowed mass region for the Higgs boson
as a function of the scale of new physics, $\Lambda$.  The region
above the solid line is forbidden because the quartic coupling,
 $\lambda(\Lambda)$,
has a Landau pole.  The region below the dot-dashed line is
forbidden because the scalar potential is unbounded from
below, $\lambda(\Lambda)< 0$. }
\end{figure} 
  
The alternate limit for the quartic coupling also gives an
interesting limit on the Higgs boson mass.  For small $\lambda$,
the evolution  with energy is
\beq
Q {d \lambda\over dQ}\sim{1\over 16\pi^2}\biggl[
B-12 g_t^4\biggr]
\quad ,
\label{smallam} 
\eeq
where
\beqn
B&=&{3\over 4}\biggl[ g^4+2 g^2 g^{\prime 2}+3 g^{\prime 4}\biggr]
\nonumber \\
g_t&=& {m_t\over v}
\quad .
\label{bgt} 
\eeqn
($g$ and $g^\prime$ are the $SU(2)_L$ and $U(1)$ coupling
constants, respectively,  and $g_t$ is the top quark Yukawa
coupling.)  
The large value of the top quark mass tends to  drive the quartic
coupling negative, thus destabilizing the 
potential.   
 Eq. \ref{smallam}  can be easily solved for $\lambda(Q)$,
\beq
\lambda(Q)\sim \lambda(\Lambda)+{B-12g_t^4\over 16 \pi^2}
\log\biggl({Q\over\Lambda}
\biggl)
\quad .
\eeq
  The requirement that the
quartic coupling be always positive  yields
a lower bound 
on the Higgs boson mass,\cite{sher} 
\beq
M_h^2 > {v^2\over 8 \pi^2}(12 g_t^4-B)\log\biggl({\Lambda\over M_h}
\biggr)
\quad .
\label{trivb} 
\eeq 
This bound is the dot-dashed curve of Fig. 1. 
Including the $2$-loop evolution of all couplings and summing
the large logarithms gives a lower bound  on the
Higgs mass slightly below 
that of Eq. \ref{trivb}.\footnote{
The lower bound on the Higgs mass is significantly weakened
if the possibility of a metastable vacuum is allowed
 for.\cite{eqpap}}
This bound is quite sensitive to the exact value of the top
quark mass.\cite{sher} 

The two bounds on the Higgs boson mass
from the scale dependence of the potential
 are shown in Fig. 1 and we see that for any given
scale $\Lambda$, the allowed region for the Higgs boson mass is
quite restricted.  For example, if there is no new physics
beyond that of the Standard Model before
the Planck scale, the allowed range is
roughly,
\beq
130~GeV < M_h < 170~GeV
,
\eeq
{\it right in the intermediate mass Higgs boson range}!
The conclusion is clear:  if the Standard Model describes
physics below $10^{16}~-~10^{18}~GeV$, then the Higgs boson must
be in the intermediate mass region.

A further indication of the importance of the intermediate mass Higgs
boson is given by fits to electroweak data.\footnote{
Indirect limits on the Higgs boson mass from electroweak data
are discussed in the chapter by A. Blondel.}  Including
data from LEP, SLD, CDF, and D0, the best fit for the
Higgs boson mass is~\cite{dem2}

\beq
M_h=149^{+148}_{-82}~GeV
\quad ,
\eeq
which gives a $95\%$ confidence level bound of 
\beq
M_h<550~GeV
\qquad .
\eeq
Since the sensitivity of the electroweak data to the Higgs boson
mass is only logarithmic, 
 the bound is unfortunately
extremely sensitive to which pieces of
data are included in the fit.
For example, removing the  measurement of
the hadronic width of the $Z$ boson  increases the
bound significantly.\cite{rosner}        The results do
seem to favor a Higgs boson in the intermediate range
however. \cite{ellis}
 
The above indications emphasize the crucial need to be able
to experimentally probe the intermediate mass region.  
 A detailed discussion of the experimental prospects for
observing an intermediate mass Higgs boson 
in many different processes is given in
Ref. \cite{gsw,gunsnow}
 and complements the discussion given here.  In 
Section 2, we discuss the Higgs boson branching ratios, including
as many radiative corrections as possible.\cite{kniehl}   
 We turn in Section 3 to a discussion of the production mechanisms at
a hadron collider and survey the
 signatures at both the LHC and the Tevatron.
More detailed discussions are contained in the chapters
by S. Mrenna and J. Gunion.  
Section 4 has an overview of the properties of an intermediate
mass Higgs boson at an $e^+e^-$ collider,\footnote{
The phenomenology of the  Standard Model
Higgs boson search at LEP and LEPII is
discussed in the chapter by P. Janot.}
 while Sections 5 and 6 summarize
the prospects for observing the intermediate mass Higgs boson 
 at a $\mu^+\mu^-$ collider and
in $\gamma\gamma$ collisions.  
Finally, Section 7 contains some conclusions.
 
\section{Higgs Branching Ratios}
\subsection{Decays to Fermion Pairs}

In the Higgs sector, the Standard Model is extremely predictive,
with all couplings, decay widths, and production cross sections
given in terms of the unknown Higgs boson mass.
  The Higgs couplings to fermions are proportional to fermion mass
and the 
gauge invariant Yukawa couplings of the Higgs boson to fermions
are,
\beq
{\cal L}_f=-\lambda_d{\overline Q}_L\phi d_R-
\lambda_u{\overline Q}_L \phi^cu_R+{\rm h.c.}
\label{yukdef}
\eeq
where $\phi^c=-i\tau_2\phi^*$ and ${\overline Q}_L=
({\overline u}_L, {\overline d}_L)$.  When the neutral Higgs
boson obtains its vev, the Yukawa couplings are fixed
in terms of the fermion masses,~\footnote{
Note that the physical Higgs boson $h$
 is given  in unitary gauge by $\phi^0=(h+v)/\sqrt{2}
$.}
\beqn
\lambda_d&=& {M_d\sqrt{2}\over v}\nonumber \\
\lambda_u&=& {M_u\sqrt{2}\over v}
\quad .
\eeqn
The measurements of the various Higgs decay channels will
therefore serve to discriminate between the Standard Model
and other models with more complicated Higgs sectors
which may have different decay chains and Yukawa
couplings.  It
is hence vital that we have reliable predictions for the
branching ratios in order to 
verify the correctness of the Yukawa couplings
of Eq.\ref{yukdef}.\cite{abdel,spirev} 

  The dominant
decays of the intermediate mass Higgs boson are into fermion-
antifermion pairs. In the Born approximation,
 the width into charged lepton pairs is 
\beq
\Gamma(h\rightarrow l^+l^-)={G_Fm_l^2\over 
4\sqrt{2}\pi} M_h \beta_l^3
\quad ,
\eeq
where
$\beta_l\equiv \sqrt{1-4m_l^2/M_h^2}$ is the velocity of the
final state leptons.  
The Higgs  boson
decay into quarks is enhanced by the color
factor $N_c=3$ and also receives significant QCD corrections,
\beq
\Gamma(h\rightarrow q {\overline q})={3G_Fm_q^2\over 4 \sqrt{2} \pi}
M_h \beta_q^3\biggl(1+{4\over 3}{\alpha_s\over \pi}\Delta_h^{QCD}
\biggr)                              
\quad ,
\eeq
where the QCD correction factor, $\Delta_h^{QCD}$,  can be found in
Ref. \cite{hbqcd}.
  
For the intermediate mass Higgs boson, 
the ${\cal O}(\alpha_s)$ corrections decrease the decay width 
for $h\rightarrow b {\overline b}$ by about a factor of $2$.  A
large portion of the corrections can be absorbed by
expressing the decay width in terms of a
running  $b$-quark mass, $m_b(\mu)$, evaluated at the
scale $\mu=M_h$.  The decay width can then be written as,\cite{dsz}
\beq
\Gamma(h\rightarrow q {\overline q})=
{3G_F\over 4 \sqrt{2} \pi} m_q^2(M_h^2)
M_h \biggl(1+5.67{\alpha_s(M_h^2)\over \pi}+\cdots
\biggr)                             , 
\eeq
where $\alpha_s(M_h^2)$ is defined in the ${\overline{MS}}$ scheme with
$5$ flavors and $\Lambda_{\overline{MS}}=150~GeV$.  
The ${\cal O}(\alpha_s^2)$ corrections are also known
in the limit $M_h>>m_q$.\cite{h2l}

In 
leading log QCD, the running of the $b$ quark mass is,
\beq
m_{b}(\mu^2)=M \biggl[{\alpha_s(M^2)
\over \alpha_s(\mu^2)}
\biggr]^{(-12/23)}
\biggl\{1+{\cal O}(\alpha_s^2)\biggr\} ,  
\label{bscale}  
\eeq
where $m_b(M^2)\equiv M$ implies that the running mass at the
position of the propagator pole is equal to the location of the pole.
For $m_b(m_b^2)=4.2~GeV$, this yields an
effective value $m_b((M_h=100~GeV)^2) = 3~GeV$.
 Inserting the QCD corrected mass into  the expression
for the width thus leads to a
significantly smaller rate than that found using $m_b=4.2~GeV$.  

The electroweak radiative corrections to $h\rightarrow q {\overline q}$
are not significant and amount to only a few 
percent correction.\cite{kniehl,ewh}
These can be neglected in comparison with the dominant
QCD corrections.  

The branching ratios for the dominant decays to fermion-
antifermion pairs are shown in Fig. 2.\footnote{
A convenient FORTRAN code for computing the QCD radiative
corrections to the Higgs boson decays is HDECAY,
which is documented in Ref. \cite{hdecay}.}  The decrease 
in the $h\rightarrow f {\overline f}$  branching ratios at $M_h\sim
150~GeV$ is due to the turn-on of the $WW^*$ decay channel.
For most of the region   below the $W^+W^-$ threshold, the Higgs
decays almost entirely to $b {\overline b}$ pairs, although it
is possible that the decays to $\tau^+\tau^-$ will
  be useful.\cite{keith}\footnote{The $\tau^+\tau^-$ decay
has large backgrounds from $t {\overline t}$ production and
from Drell-Yan production of $\tau^+\tau^-$ pairs which may
be insurmountable.}    
The other fermionic Higgs boson 
decay channels are almost certainly too small to be
separated from the backgrounds.
Even including the QCD corrections, the rates
can be seen to roughly
scale with the fermion masses and the color factor, $N_c$,
\beq
{\Gamma(h\rightarrow b {\overline b})\over
\Gamma(h\rightarrow \tau^+\tau^-)}\sim
{3 m_b(M_h^2)^2\over m_\tau^2},
\eeq  
and so a measurement of the branching ratios could
serve to verify the correctness of the Standard Model
couplings. 
The largest uncertainty is in the value of $\alpha_s$, which
affects the running $b$ quark mass, as in Eq. \ref{bscale} .

\begin{figure}[tb]
\centerline{\epsfig{file=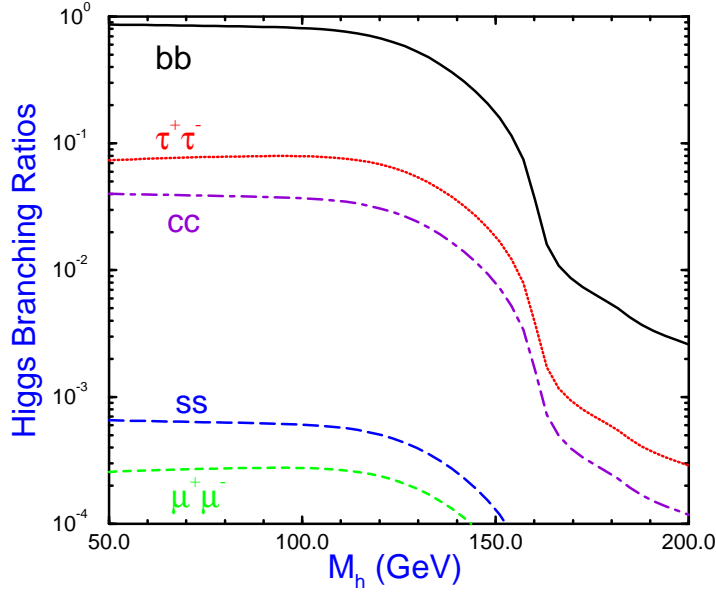,height=3.5in}}
\caption{Branching ratios of the Standard Model Higgs boson
 to fermion-antifermion pairs, including 
QCD radiative corrections. We have taken $M_t=175~GeV$.
The radiative corrections were computed using the program 
HDECAY.$^{22}$}
\end{figure}

\subsection{Decays to Gauge Boson Pairs}

The decay of the Higgs boson to gluons arises through
fermion loops,\cite{hhg}
\beq
\Gamma_0(h\rightarrow gg)={ G_F \alpha_s^2 M_h^3
\over 64 \sqrt{2}\pi^3}
\mid \sum_q F_{1/2}(\tau_q)\mid^2
\eeq
where $\tau_q\equiv
4 m_q^2/M_h^2$ and  $F_{1/2}(\tau_q)$ is defined,
\beq
F_{1/2}(\tau_q) \equiv -2\tau_q\biggl[1+(1-\tau_q)f(\tau_q)\biggr]
\quad .
\label{etadef}
\eeq
The function $f(\tau_q)$ is given by,
\beq
f(\tau_q)=\left\{\begin{array}{ll}
\biggl[\sin^{-1}\biggl(\sqrt{1/\tau_q}\biggr)\biggr]^2,&\hbox{if~} 
\tau_q\ge 1
\\
-{1\over 4}\biggl[\log\biggl({x_+\over x_-}\biggr)
-i\pi\biggr]^2,
&\hbox{ if~}\tau_q<1,
\end{array} 
\right .
\label{fundef}
\eeq
with
\beq
x_{\pm}=1\pm\sqrt{1-\tau_q}
.
\eeq
In the limit in which the quark mass is much less than the Higgs boson mass,
(the relevant limit for the $b$ quark),
\beq
F_{1/2}\rightarrow {2 m_q^2\over M_h^2}\log^2\biggl(
{m_q\over M_h}\biggr)
.
\eeq 
On
the other hand, for a heavy quark, $\tau_q\rightarrow\infty$,
 and $F_{1/2}(\tau_q)$ approaches
a constant,
\beq
F_{1/2}\rightarrow -{4\over 3}
.
\label{f12}
\eeq
It is clear that the dominant contribution to the gluonic decay of the 
Higgs boson is from the top quark loop and from possible new generations of
heavy fermions.   A measurement of this rate would serve to count
the number of heavy fermions since the heavy fermions do not
decouple from the theory.

The QCD radiative corrections
from  $h\rightarrow ggg$ and $h\rightarrow
g q {\overline q}$
to the hadronic decay of the Higgs boson
are large since they typically increase the
width by more than $50\%$.  The radiatively corrected width can be
approximated by
\beq
\Gamma(h\rightarrow ggX)=
\Gamma_0(h\rightarrow gg)\biggl[
1+E{\alpha_s(\mu)\over \pi}\biggr]
\quad ,
\eeq
where $E={215\over 12}-
{23\over 6}\log(\mu^2/M_h^2)$.\cite{dsz,hggg}
The radiatively corrected branching ratio for $h\rightarrow
ggX$ is the solid curve in Fig. 3. 

The intermediate mass Higgs boson can also decay to vector
boson pairs $V V^*$, ($V=W^\pm,Z$),
 with one of the gauge bosons off-shell. 
The widths, summed over all available channels for $V^*\rightarrow
f {\overline f}$ are: ~\cite{wkwm}
\beqn
\Gamma(h\rightarrow Z Z^*)&=&
{g^4 M_h\over 2048 (1-x_W)^2\pi^3}
\biggl(7-{40\over 3}x_W+{160\over 9}x_W^2
\biggr)
  F\biggl({M_Z\over M_h}\biggr)
\nonumber \\
\Gamma(h\rightarrow W W^*)&=&{3 g^4 M_h\over 512 \pi^3} F\biggl({M_W\over M_h}
\biggr)
\eeqn
where $x_W\equiv \sin^2\theta_W$ and 
\beqn
F(x)&\equiv & -\mid 1-x^2\mid
\biggl({47\over 2} x^2-{13\over 2} +{1\over x^2}\biggr)
-3\biggl(1-6 x^2+4 x^4\biggr)
\mid \ln (x) \mid 
\nonumber \\  && 
+3{1-8 x^2+20 x^4\over \sqrt{4 x^2-1}}
\cos ^{-1}\biggl( {3 x^2-1\over 2 x^3}\biggr)
\quad .
\eeqn
These widths can be significant when the Higgs boson mass
approaches the real $W^+W^-$ and $ZZ$ thresholds, as can be seen
in Fig. 3. The $WW^*$ and $ZZ^*$ branching ratios grow
rapidly with increasing Higgs mass and above $2M_W$,
the rate for $h\rightarrow W^+W^-$ is close to 1.   
The decay to $ZZ^*$ is roughly an order of magnitude
smaller than the decay to $WW^*$ over
much of the intermediate mass Higgs range due to the smallness
of the neutral current couplings as compared to the
charged current couplings.

\begin{figure}[tb]
\centerline{\epsfig{file=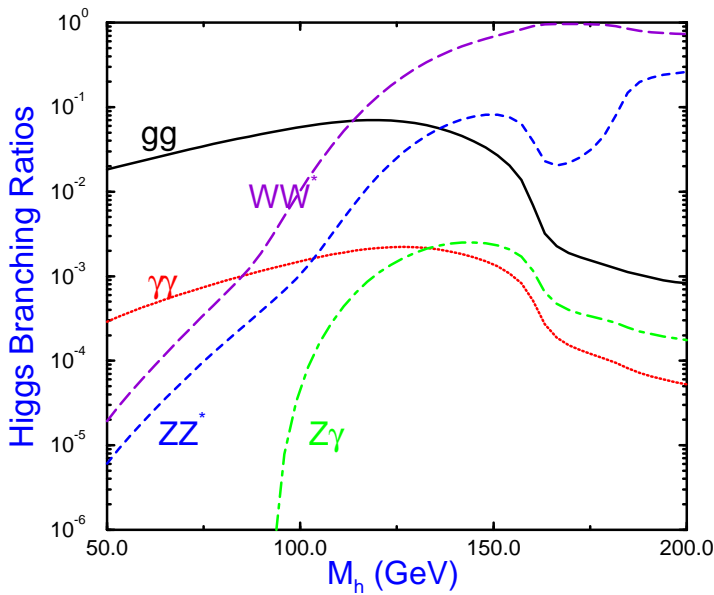,height=3.5in}}
\caption{Branching ratios  of the Standard Model Higgs
boson to gauge boson pairs.  The rates to $WW^*$ and $ZZ^*$ must be
multiplied by the appropriate branching ratios for $W^*$ and $Z^*$
decays to $f {\overline f} $ pairs. We have taken $M_t=175~GeV$.
The radiative corrections were computed using the program 
HDECAY.$^{22}$}
\end{figure}

The decay $h\rightarrow Z \gamma$ is not useful phenomenologically,
so we will not discuss it here although
  the complete expression for the
branching ratio can be found in Ref. \cite{hzg}. 
On the other hand, the decay $h\rightarrow \gamma \gamma$ is an
important mode for the Higgs search at the LHC. 
At lowest order, the branching
ratio is, \cite{hgg} 
\beq
\Gamma(h\rightarrow \gamma\gamma)={\alpha^2 G_F\over 128\sqrt{2} \pi^3}
M_h^3\mid \sum_i N_{ci} Q_i^2 F_i(\tau_i)\mid^2 
\eeq
where the sum is over fermions and  $W^\pm$ bosons with
$F_{1/2}(\tau_q)$ given in Eq. 23, and 
\beq
F_W(\tau_W)= 2+3\tau_W[1+(2-\tau_W)f(\tau_W)]
\quad .
\label{fdef}
\eeq 
$\tau_W=4 M_W^2/M_h^2$, $N_{Ci}=3$ for quarks
and $1$ otherwise, and $Q_i$ is the electric charge in units of $e$.
The function $f(\tau_q)$ is given in Eq. \ref{fundef}. 
The $h\rightarrow \gamma\gamma$ decay channel clearly
probes the possible existence of heavy charged particles.  
(Charged scalars, such as those existing in SUSY models,
would also contribute to the rate.)\cite{hhg}

\begin{figure}[tb]
\centerline{\epsfig{file=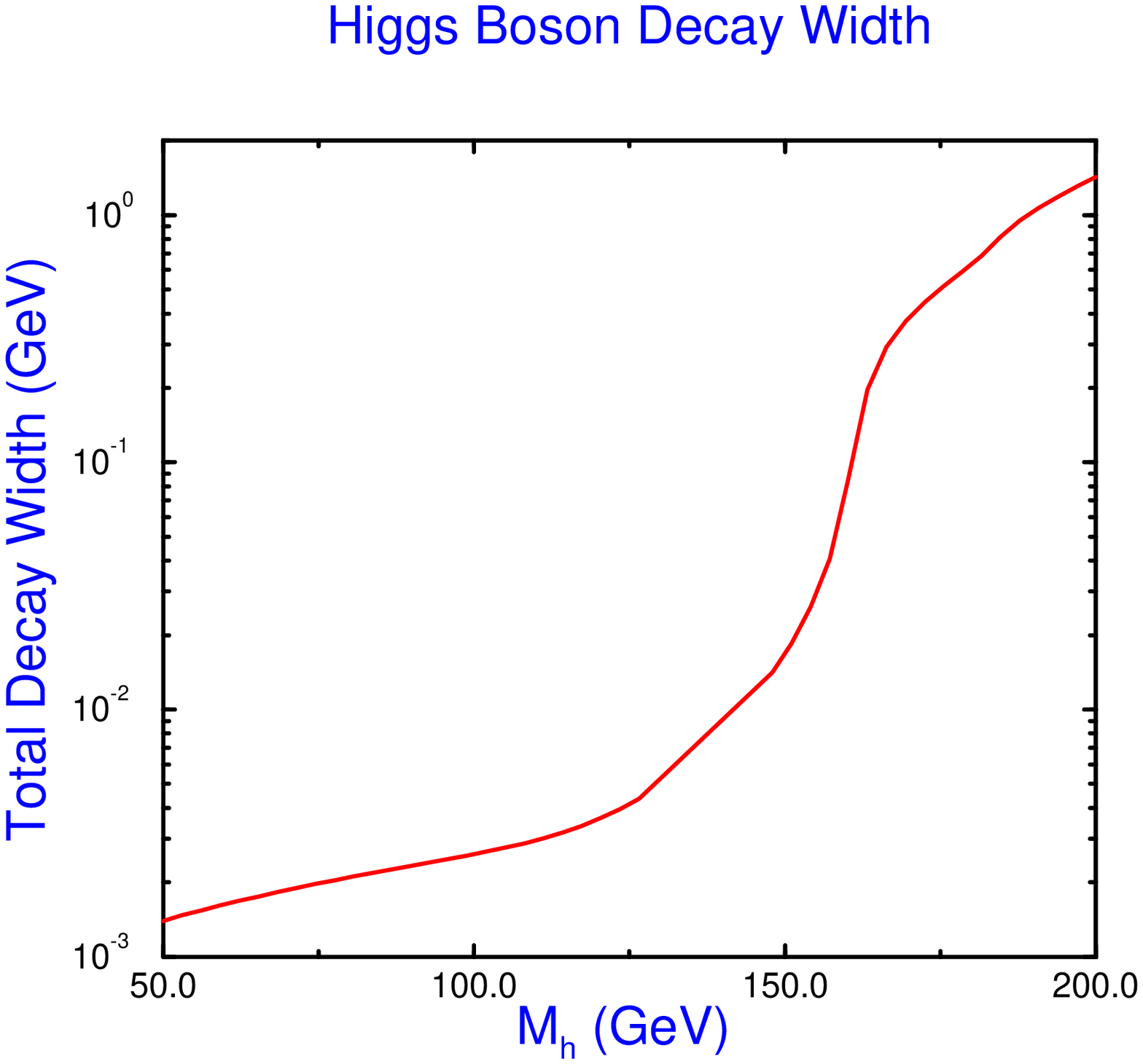,height=3.5in}}
\caption{Total Higgs boson decay width in the Standard Model,
including QCD radiative corrections. The turn-on of 
the $W^+W^-$ threshold 
at $M_h\sim 160~GeV$ is obvious. The top quark mass is fixed
to $M_t=175~GeV$.}
\end{figure}  

In the limit where the particle in the loop is much heavier 
than the Higgs boson, $\tau\rightarrow \infty$, 
\beqn
F_{1/2}&\rightarrow & -{4\over 3}
\nonumber \\
F_W&\rightarrow & ~7
\quad .
\label{fwdef}
\eeqn
The top quark contribution $(F_{1/2})$ is therefore much
smaller than that of the $W$ loops $(F_W)$ and so we expect the QCD
corrections to be 
less important than is the case for the $h\rightarrow gg$
decay. 
In fact the QCD corrections to the total width for $h\rightarrow
\gamma\gamma$ are quite small.\cite{spgg} 
The $h\rightarrow \gamma\gamma$ branching ratio is
the dotted line in Fig. 3.
For small Higgs masses it rises with increasing $M_h$
and peaks at around $2\times 10^{-3}$ for $M_h\sim 125~GeV$.  
Above this mass, the $W W^*$ and $ZZ^*$ decay modes are
increasing rapidly with increasing Higgs mass and the 
$\gamma \gamma$ mode
becomes tiny.

The total width for the intermediate mass Higgs boson is shown in
Fig. 4.  Below around $M_h\sim 150~GeV$, the Higgs boson is
quite
narrow with $\Gamma_h< 10~MeV$.  As the $WW^*$ and $ZZ^*$ channels
become accessible, the width increases rapidly with $\Gamma_h\sim 1~GeV$
at $M_h\sim 200~GeV$.  In the intermediate mass region,
the Higgs boson width is too narrow to be resolved experimentally.
The total width for the lightest neutral Higgs boson in
the minimal SUSY model is typically much smaller than the 
Standard Model width for the same  Higgs boson 
mass and so a measurement
of the total width could serve to discriminate between the
two models.
The individual decay channels are also quite different in the
Standard Model and in a SUSY model, as is discussed in the
chapter by H.~Haber.

We turn now to the production of the intermediate mass Higgs
boson in hadronic interactions.

\section{Higgs Production in Hadronic Interactions}
\subsection{Gluon Fusion}
\subsubsection{Lowest Order}

An important production mechanism for the Higgs boson at
hadron colliders is gluon fusion which proceeds through 
a quark triangle.
This is 
the dominant contribution to Higgs boson production at the
LHC for all $M_h<1~TeV$. 
The lowest order cross section
for $gg\rightarrow h$ is,\cite{hhg}
\beqn
{\hat \sigma}(gg\rightarrow h)&= &
{\alpha_s^2\over 1024 \pi v^2}
\mid \sum_q  F_{1/2}(\tau_q)\mid^2\delta    
\biggl(1-{{\hat s}\over M_h^2}\biggr)
\nonumber \\
&\equiv &{\hat \sigma}_0(gg\rightarrow h)\delta
\biggl(1-{{\hat s}\over M_h^2}\biggr)
\quad ,
\label{sigdef}
\eeqn     
where ${\hat s}$ is the gluon-gluon center of mass energy,
$v=246~GeV$
and $F_{1/2}(\tau_q)$ is defined in Eq. \ref{etadef}.
In the heavy quark
limit, $(M_t/M_h) \rightarrow \infty$, the cross
section becomes,
\beq
{\hat \sigma}_0(gg\rightarrow h)\sim {\alpha_s^2\over
576 \pi v^2}      \quad .
\eeq
This rate counts the number of heavy quarks and  so could
be a window into a possible fourth generation of quarks.

The Higgs boson production cross section at a hadron collider can be 
found by integrating the parton cross section with the gluon
structure functions, $g(x)$,
\beq
\sigma_0(pp\rightarrow h)={\hat \sigma_0}\tau\int_\tau^1
{dx\over x} g(x)g\biggl({\tau\over x}\biggr),
\label{logh}
\eeq
where $\sigma_0$ is given in Eq. \ref{sigdef},
 $\tau\equiv M_h^2/S$, and $S$ is
the hadronic center of mass energy.   

\begin{figure}[tb]
\epsfig{file=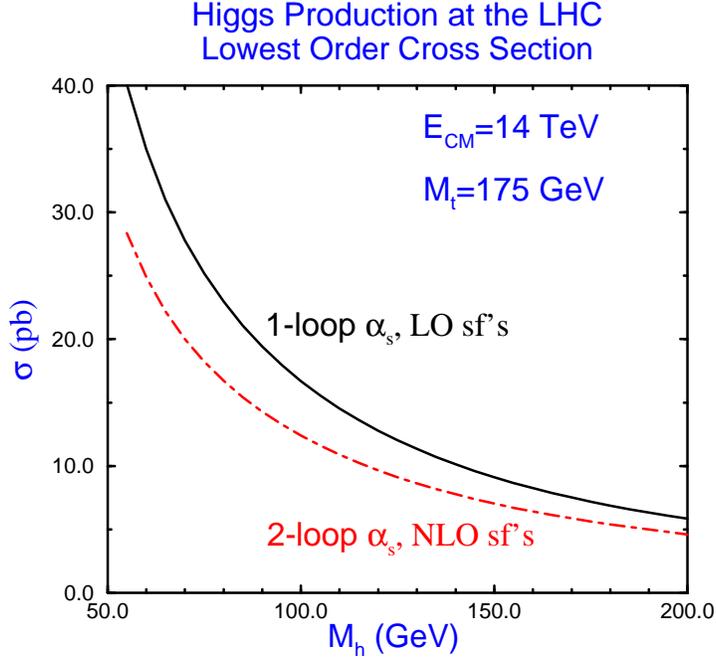,height=3.5in}
\caption{Cross section for gluon fusion, $gg\rightarrow h$, at the
LHC using the lowest order parton cross section of Eq. 34.  The
solid (dot-dashed) line uses the one-loop (two-loop) expression
for $\alpha_s(\mu)$, along with structure functions fitted
to lowest order (next to lowest order) expressions for the 
data.  The renormalization scale is $\mu=M_h$.}
\end{figure}

We show the rate obtained using the lowest order parton cross section
of Eq. \ref{sigdef}
in Fig. 5 for the LHC.  When computing the lowest order result 
from Eq. \ref{sigdef}, it is ambiguous whether to use the one-
 or  two- loop
equations for $\alpha_s(\mu)$
 and which structure functions to use;
a set fit to data using only the lowest order  in $\alpha_s$ predictions
or a set which includes some higher order effects.  The difference
between the equations for $\alpha_s$ and the different structure functions
is ${\cal O}(\alpha_s^2)$ and hence  higher order in $\alpha_s$ when one is
computing the ``lowest order'' result.  In Fig. 5, we show two 
different definitions of the lowest order result and see that they
differ significantly from each other.  We will see in the next
section that the result obtained using the $2$-loop $\alpha_s$ and
NLO structure functions, but the lowest order parton  cross section,
is a poor approximation to the radiatively corrected rate.  
Fig. 5 takes the scale factor $\mu=M_h$ and  the results are
quite sensitive to this choice.  

\subsection{QCD Corrections to $gg\rightarrow h$}

In order to obtain reliable predictions
for the production rate, it is important to compute
the $2-$loop QCD
radiative corrections to $gg\rightarrow h$.  The complete
${\cal O}(\alpha_s^3)$
calculation is available in Refs. \cite{spirev,hqcd}.
The analytic result is quite complicated, but the 
computer code including all QCD radiative corrections
is readily available.\cite{hdecay}  

For the intermediate
mass Higgs boson, the result in the $M_h/M_t\rightarrow 0$ limit
turns out to be an excellent approximation to the
exact result and can be used in most cases.  The 
heavy quark limit can be obtained from
 the gauge invariant effective Lagrangian,\cite{daw}
\beq
{\cal L}=-{1\over 4} \biggl[1-{2\beta_s\over g_s(1+\delta)}
{h\over v}\biggr]
G_{\mu\nu}^AG^{A\mu\nu}
-{M_t\over v} {\overline t} t h
\quad ,
\label{lowen}
\eeq
where $\delta=2\alpha_s/\pi$ is the
anomalous mass dimension  arising from the renormalization
of the $t {\overline t}h$ Yukawa coupling constant, $g_s$ is the
QCD coupling constant, and $G_{\mu\nu}^A$ is the color $SU(3)$
field.  This Lagrangian can be derived using low energy
theorems which are valid in the limit $M_h<<M_t$
and  yields momentum dependent $ggh$, $gggh$,
and $ggggh$ vertices which can be used to compute the 
rate for $gg\rightarrow h$ to ${\cal O}(\alpha_s^3)$.  
  
Since the $hgg$ coupling  in the $M_t\rightarrow\infty $ limit
results from heavy fermion loops, it is
only the heavy fermions which contribute to $\beta_s$ in Eq.
\ref{lowen}.  To 
${\cal O}(\alpha_s^2)$, the heavy fermion contribution to the QCD $\beta$ 
function is,
\beq
{\beta_s\over g_s}\mid{\rm heavy~fermions}=
N_H{\alpha_s\over 6\pi}\biggl[
1+{19\alpha_s\over 4}\biggr]
\quad .
\eeq
 $N_H$ is the number of heavy fermions.

The parton level cross section for $gg\rightarrow h$ is
found by computing the ${\cal O}(\alpha_s^3)$ virtual
graphs for $gg\rightarrow h$ and combining them with the 
bremsstrahlung process $gg\rightarrow gh$.  The answer
in the heavy top quark limit is,\cite{spirev,hqcd,daw}
\beq 
{\hat \sigma}_1(gg\rightarrow h X)
={\alpha_s^2(\mu)\over 576 \pi v^2}
\biggl\{ \delta(1-z)+{\alpha_s(\mu)\over\pi}
\biggl[ h(z)+{\bar h}(z)\log\biggl({M_h^2\over \mu^2}\biggr)
\biggr]\biggr\}
\eeq
where
\beqn
h(z)&=& \delta(1-z)\biggl(\pi^2+{11\over 2}\biggr)-{11\over 2}(1-z)^3
\nonumber \\
&&+6\biggl(1+z^4+(1-z)^4\biggr)\biggl({\log(1-z)\over 1-z}\biggr)_+
-{\overline h}(z) \log(z)
\nonumber \\
{\bar h}(z)&=& 6 \biggl({z^2\over (1-z)_+} 
+(1-z)+z^2(1-z)\biggr)
\eeqn
and $z\equiv M_h^2/{\hat s}$. 
The answer is written in terms of ``+'' distributions,
which are defined by the integrals,
\beq
\int_0^1{f(x)\over (1-x)_+}\equiv\int^1_0{f(x)-f(1)
\over 1-x}
\quad .
\eeq
The factor $\mu$ is an arbitrary renormalization
point.  To $\alpha_s^3$, the physical hadronic cross
section is independent of $\mu$.  
  There are also ${\cal O}
(\alpha_s^3)$ contributions from $q {\overline q}$, $qg$
and ${\overline q}g$
initial states, but these are numerically small.

\begin{figure}[tb]
\centerline{\epsfig{file=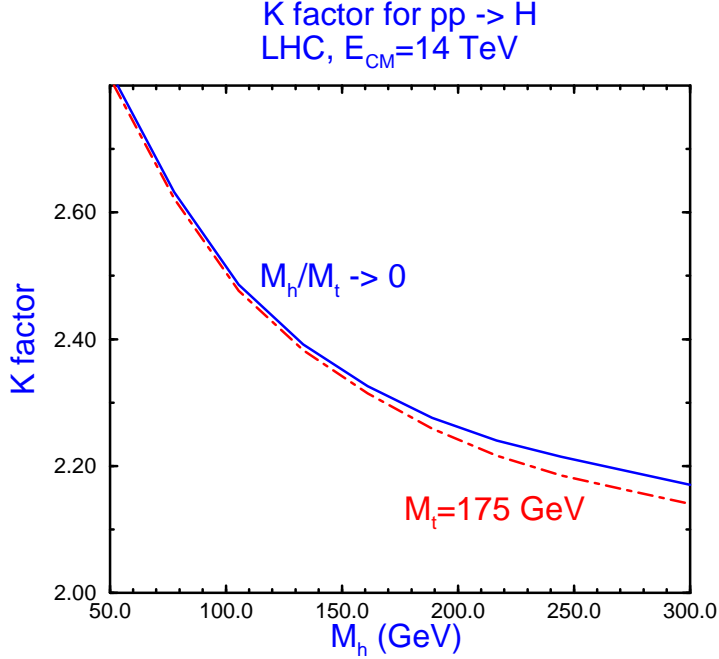,height=3.5in}}
\caption{K factor as defined in Eq. 42.  We have
taken the renormalization scale $\mu=M_h$. The solid curve
is the result obtained using the $M_t\rightarrow \infty$ limit,
Eqs. 35 and 39, and the dot-dashed curve is the exact result from
Ref. 17.}
\end{figure} 
We can define a $K$ factor as
\beq
K\equiv{{\sigma}_1(pp\rightarrow hX)\over
{ \sigma}_0(pp\rightarrow h)}
\quad ,
\eeq
where $\sigma_1(pp\rightarrow hX)$ is the ${\cal O}(\alpha_s^3)$
radiatively corrected rate for Higgs production and $\sigma_0$
is the lowest order rate found from in Eq. \ref{sigdef}.
From Eq. 40, it is apparent that a significant portion of the corrections
result from the rescaling of the lowest order result,
\beq
K\sim 1+{\alpha_s(\mu)\over \pi}\biggl[ \pi^2+{11\over 2}
+ ...\biggr]
\quad .
\eeq
Of  course $K$ is not a constant, but depends on the renormalization
scale $\mu$ as well as $M_h$.  In Fig. 6, we show the $K$ factor obtained
in the limit $M_h/M_t\rightarrow 0$,
as well as the exact $K$ factor found in Ref. \cite{spirev}. 
The radiatively corrected cross section ${\hat \sigma}_1$ should
be convoluted with next-to-leading order structure functions,
while it is ambiguous which structure
functions and definition of $\alpha_s$ to use
in defining the lowest order result, ${\hat \sigma}_0$, as
discussed above.  Fig. 6 uses the $2$-loop definition of
$\alpha_s$ and NLO structure functions for both $\sigma_0$
and $\sigma_1$.  A $K$ factor computed using lowest order 
structure functions and the one-loop definition of $\alpha_s$
for $\sigma_0$ would be smaller than that shown in Fig. 6,
(as is obvious from Fig. 5).

The
$K$ factor varies between $2$ and $3$ for the intermediate mass
Higgs boson and so the QCD corrections significantly increase the
rate from the lowest order result.
It is evident from Fig. 6 that
for the intermediate mass Higgs boson, the heavy top quark limit is an
excellent approximation to the $K$ factor.  
The easiest way to compute the radiatively corrected cross section is
therefore to calculate the lowest order cross section including the
complete mass dependence of Eq. \ref{sigdef}
 and then to multiply by the $K$ factor
computed in the $M_t\rightarrow \infty$ limit.
This result will be extremely accurate.

The other potentially
 important correction to the $hgg$ coupling is the
two-loop electroweak
 contribution involving the top quark, which is of
${\cal O}(\alpha_S G_F M_t^2)$.  In the heavy quark limit,
the function $F_{1/2}(\tau_q)$
 of Eq. \ref{etadef} receives a contribution,~\cite{kniehl,dg}
\beq
F_{1/2}(\tau_q)\rightarrow
F_{1/2}(\tau_q)\biggl(1+{G_FM_t^2\over 16\sqrt{2}\pi^2}
\biggr)
\quad .
\eeq
When the total rate for Higgs production is computed, the
${\cal O}(\alpha_s G_F M_t^2)$ contribution is $<.2\%$ and
so can be neglected.  
The ${\cal O}(\alpha_s G_F M_t^2)$ contributions 
therefore do not spoil
the usefulness of the $gg\rightarrow h$ mechanism 
as a means of  counting
heavy quarks.

\begin{figure} [tb] 
\centerline{\epsfig{file=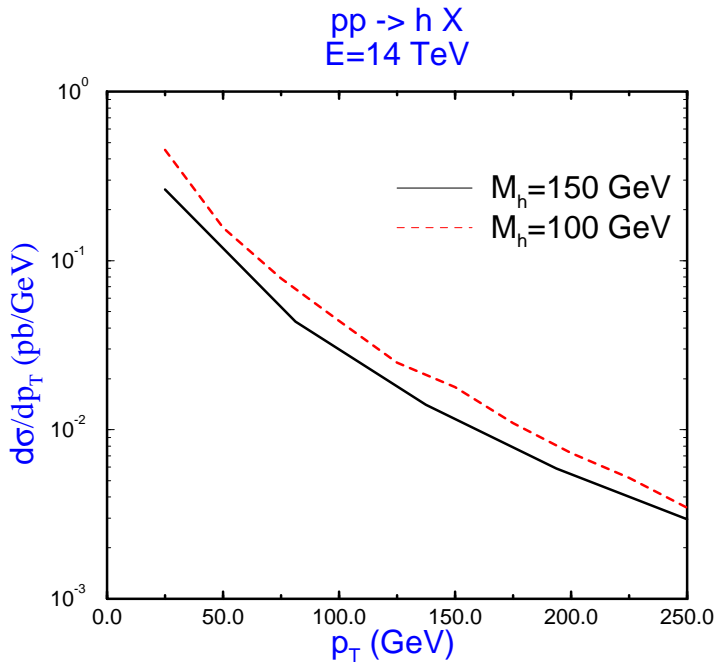,height=3.5in}}
\caption{Higgs boson transverse momentum distribution 
 from Eq. 45.}
\end{figure}

At lowest order the gluon fusion process yields a Higgs boson with
no transverse momentum. 
At the next order in perturbation theory, gluon fusion
produces a Higgs boson with finite $p_T$, primarily through
the process $gg\rightarrow gh$.  As $p_t\rightarrow 0$, 
the parton cross section diverges as $1/p_T^2$,\cite{qcdsum}
\beqn 
{d{\hat\sigma}\over d {\hat t}}(gg\rightarrow g h)
&=&{\hat \sigma}_0{3 \alpha_s\over 2 \pi}\biggl\{
{1\over p_T^2}\biggl[
\biggl(1-{M_h^2\over {\hat s}}\biggr)^4 + 1
+\biggl( {M_h^2\over {\hat s}}\biggr)^4\biggr]
\nonumber \\
&& - {4\over {\hat s}}\biggl(1-{M_h^2\over {\hat s}}
\biggr)^2+{2 p_T^2\over {\hat s}}\biggr\}\quad .
\label{pth}
\eeqn 
 The hadronic cross section can be found by integrating
 Eq. \ref{pth} with
the gluon structure functions.  In Fig. 7, we show the $p_T$
spectrum of the Higgs boson at ${\cal O}(\alpha_s^3)$.
The event rate even at large $p_T$ is significant.  This figure
clearly demonstrates the singularity at $p_T=0$.

The terms which are singular as $p_T\rightarrow 0$ can 
be isolated and the integrals performed explicitly.
For simplicity we consider only the $gg$
initial state.
\beq
{d \sigma\over d p_T^2 d y} (pp\rightarrow g h)\mid _{
p_T^2\rightarrow 0}\sim
{\hat \sigma}_0 { 3\alpha_s\over 2 \pi} {1\over p_T^2}
\biggl[
6 \log \biggl({M_h^2\over p_T^2}\biggr) - 2 \beta_0
\biggr]
g(\tau e^y)g (\tau e^{-y})+...
\eeq
where $\tau=M_h^2/S$, $\beta_0=(33-2 N_L)/6$, and $N_L=5$ is
the number of light flavors.  Clearly when $p_T << M_h$, the terms 
containing the large logarithms resulting from the multiple gluon emission 
must be summed.  A consistent procedure for resumming the
logarithms to next to leading order has been found
by Collins, Soper, and Sterman. \cite {colsop}  At an intermediate value
of $p_T$, one must then switch from the resummed result to the
perturbative result which is valid at high $p_T$.
This results in a flattening of the $d\sigma/dp_T$
curve of Fig. 7 at low $p_t$.\cite{qcdsum,soft} 

\begin{figure}[htb]
\centerline{\epsfig{file=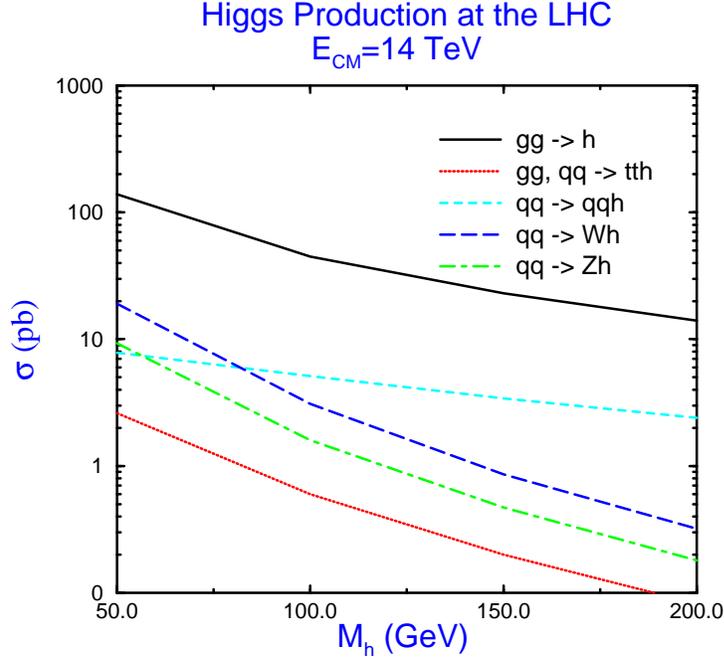,height=3.5in}}
\caption{Contributions to $pp\rightarrow h X$ at the LHC,
$E_{CM}=14~TeV$, including QCD radiative corrections for all
processes except $t {\overline t} h$ production.}
\end{figure}

 Fig. 8 shows the various contributions to Higgs boson production in
the intermediate mass region at the LHC,
including QCD corrections for all processes except
$t {\overline t} h$ production.\footnote{A discussion
of the uncertainties in the calculations of the Standard
Model Higgs boson production rates at the LHC is
given in Ref. \cite{stirl}.}  Gluon fusion 
(the solid curve) is obviously
the dominant mechanism with a cross
section between $10$ and $100~pb$
over the entire intermediate mass region.
Vector boson fusion, $qq\rightarrow qqh$, (the short-dashed
line) is an order of magnitude smaller than gluon fusion
for the intermediate mass Higgs boson.  Since it is not
useful for Higgs searches in this region, we will not
discuss it.

    There are two particularly important decay modes
for the experimental searches in the
intermediate mass region at the LHC.
For $M_h<  125~GeV$, the best signal is from $h\rightarrow \gamma
\gamma$.  The branching ratio is larger than $ 10^{-3}$ for
$80~GeV < M_h < 120~GeV$ and so $100~fb^{-1}$ will produce
$1000$'s of $h\rightarrow \gamma\gamma$ events.
The Higgs signal is then a narrow bump in the $\gamma \gamma$
invariant mass spectrum. 
There is a very large irreducible
background from two photon events, as well
as a significant background
 from jets which are misidentified as photons.
 With $100~fb^{-1}$, ATLAS
expects to probe $110~GeV < M_h < 140~GeV$, while CMS 
estimates that it will be sensitive
to $85~GeV < M_h < 150~GeV$.\cite{atcms}
The CMS detector will cover a larger range in Higgs mass, due to the
expected excellence of its electromagnetic calorimeter.
 For the lower Higgs boson masses, the rate
to $\gamma \gamma$ becomes extremely small
(see Fig. 3) and the backgrounds become very large.
One must hope that the unexplored region below $85~GeV$ will be
probed at LEPII.

The $\gamma\gamma$ mode will also allow a precise measurement of
the Higgs mass.  For $M_h\sim 100~GeV$, ATLAS expects to
measure
$\Delta M_h\sim 1.4~GeV$, while CMS claims a slightly
better precision. \cite{atcms}\footnote{
By combining various channels and the results
from both detectors, Ref. \cite{gunsnow}
finds that for $M_h=100~GeV$, a
measurement of $\Delta M_h\sim 96~GeV$ will
be obtainable at the LHC with a luminosity
of $L=300~fb^{-1}$.}
 A more complete discussion
of the physics capabilities of ATLAS and CMS to detect the
Higgs boson in the $\gamma
\gamma$ decay mode  can be found in the chapter by J.~ Gunion. 

 Above  $M_h \sim 125~GeV$, the branching ratio
 to $h\rightarrow \gamma\gamma$ 
becomes too small to be useful and the best signal is from
$h\rightarrow Z Z^*\rightarrow l^+l^- l^{\prime +} l^{\prime -}$.
Above the $ZZ$ threshold, this is termed the ``gold-plated''
mode because it is so easy to observe.  
\cite{gwud} Below the $ZZ$ threshold,
the irreducible backgrounds from
$ZZ^*$ and $Z\gamma^*$ production are small and the
dominant reducible backgrounds are from $t {\overline t}$
and $Z b {\overline b}$ production which can be efficiently
eliminated with lepton isolation cuts.
With $100~fb^{-1}$, both detectors expect that the smallest
Higgs mass which they will be  sensitive to in the $ZZ^*$ channel
will be $M_h\sim 130~GeV$, while they
will probe masses up to $M_h\sim 400-500~GeV$
in the $4$ charged lepton channel.\cite{atcms}
  To probe down to $M_h\sim 120~GeV$
in this mode 
will require several years of running at high luminosity. 
The ATLAS
and CMS technical proposals contain numerous details.\cite{atcms}

A measurement of both the $h\rightarrow \gamma\gamma$ and
$h\rightarrow Z Z^*$ decay modes would be an important 
test of the $SU(2)$ symmetry of the Standard Model.
  The $h\rightarrow \gamma\gamma$
mode is dominated by the $W$ boson
loop (see Eq. \ref{fwdef}) and so is
sensitive to the $hW^+W^-$ coupling, while the $h\rightarrow
ZZ^*$ mode probes the $hZZ$ coupling.  In the Standard Model,
these couplings are related by,
\beq
{g_{hWW}\over g_{hZZZ}}=\cos^2\theta_W
\quad , 
\eeq
while the ratio of couplings is quite different in a supersymmetric
model.  

\subsection{Associated Production, $pp(p {\overline p}
)\rightarrow V h$.}

The process $q {\overline q}\rightarrow Wh$ offers the hope 
of being able to tag the Higgs boson by the 
$W$ boson decay products.\cite{stange} 
This process has the rate:
\beq
{\hat \sigma}(q_i {\overline q}_j\rightarrow W^\pm h)
={G_F^2 M_W^6 \mid V_{ij}\mid ^2\over 6 {\hat s}^2
(1-M_W^2/{\hat s})^2}\lambda_{Wh}^{1/2}\biggl[
1+{{\hat s}\lambda_{Wh}\over 12 M_W^2}\biggr]
\eeq
where $\lambda_{Wh}
=1-2(M_W^2+M_h^2)/{\hat s}+(M_W^2-M_h^2)^2/{\hat s}^2$
and $V_{ij}$ is the Kobayashi-Maskawa angle associated with the
$q_i {\overline q}_j W$ vertex.  
This process is sensitive to the $W^+W^-h$ coupling and
so will be different in extensions of the Standard Model.

\begin{figure}[tb] 
\centerline{\psfig{file=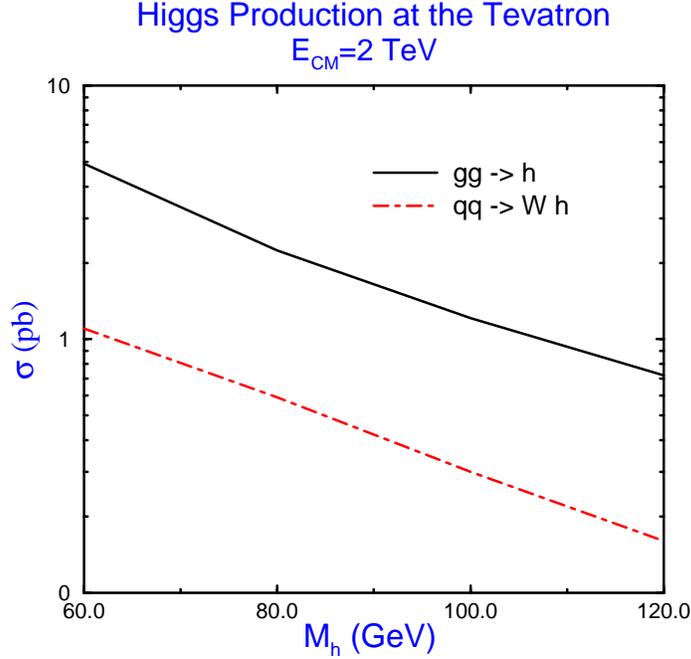,height=3.5in}}
\caption{Next to leading order QCD  predictions for Higgs
boson production at the Tevatron.
 The dot-dashed line is the $W^\pm h$ production rate (summed
over $W^\pm$ charges),
while the solid line is the rate for
Higgs production from gluon
fusion.}
\end{figure}  

Since this mechanism produces a relatively small number of
signal events, (as can be seen clearly in Figs. 8 and 9),
it is
important to compute the rate as accurately as possible
by including the QCD radiative corrections.  This has been
done in Ref. \cite{hw}, where  it is shown that
the cross section  can be  written as
\beq
{d\sigma\over d q^2
} (pp\rightarrow W^\pm h)=\sigma (pp\rightarrow W^{\pm *}
) 
{G_F M_W^4\over \sqrt{2}\pi^2 (q^2-M_W^2)^2}
{\mid {\vec p}\mid\over \sqrt{q^2}}\biggl(
1+{\mid {\vec p}\mid^2\over 3 M_W^2}\biggr)
\label{wvh}
\eeq 
to all orders in $\alpha_s$.  In Eq. \ref{wvh}, $W^*$ is a virtual
$W$ with momentum $q$ and $\mid {\vec p}\mid= 
\sqrt{s}\lambda_{Wh}^{1/2}/2$ is
the momentum of the outgoing $W^\pm$ and $h$.
  From Eq. \ref{wvh}, it is clear
that the radiative corrections to $W^\pm h$ production are identical to
those for the Drell-Yan process which have been known
for some time.\cite{dy}  Using the DIS factorization scheme, the
cross section at the LHC is increased by roughly $17\%$ 
over the lowest order rate.  The QCD
corrected cross section is relatively insensitive to the choice
of renormalization and factorization scales.  It is, however,
quite sensitive to the choice of structure functions.  The rate for
$pp\rightarrow W^\pm h$ at the LHC is shown in Fig. 8
(the long-dashed curve) and is more than
an order of magnitude smaller than the rate from gluon fusion.
The rate for $pp\rightarrow Zh$ is smaller still.  
   
The $Wh$ events can be tagged by identifying the charged lepton from the
$W$ decay.  Imposing isolation cuts on the lepton significantly reduces the
background.  At the LHC, there are sufficient events that the Higgs
produced in association with a $W$ 
boson can be identified through the $\gamma\gamma$ decay mode.\cite{stange}
ATLAS claims a $4\sigma$ signal in this channel for $80~GeV < M_h < 120~GeV$
with $100 fb^{-1}$, (this corresponds to about $15$ signal events),
while CMS hopes to find a $6-7\sigma$ effect in this channel.
There  are a large number of $Wh$ events with $W\rightarrow l \nu$
and $h\rightarrow b {\overline b}$, but unfortunately the
backgrounds  to this decay chain
are difficult to reject and observation of this signal
will probably require a high luminosity, $L=10^{33}/cm^2/sec$.\cite{stange}

Associated production of a Higgs boson with a $W^\pm$ boson can also
potentially be observed at the Tevatron.\cite{tev2000}
For a $100~GeV$ Higgs boson, the lowest order cross section
is $.2~pb$.  Including the next-to-leading order corrections
increases this to $.3~pb$, while summing over the soft
gluon effects increases the NLO result by $2-3\%$.
\cite{mren}
The next to leading order rate is shown in Fig. 9 and is much
smaller that that from gluon fusion.\footnote{I appears
hopeless to search for the Higgs boson from $gg\rightarrow h$
at the Tevatron.}  
  At the
Tevatron, the Higgs boson  from $W^\pm h$
production must be searched for in the $b {\overline b}$
decay mode since the $\gamma\gamma$ decay mode produces
too few events to be observable.
  The largest backgrounds are $W b {\overline b}$ 
and $WZ$, along with top quark production.
The background from top quark production is considerable smaller at
the Tevatron than at the LHC, however. 
  The Tevatron with 
$2-4~fb^{-1}$ will be sensitive to $M_h< M_Z$, a region which
presumably will already be probed by LEPII.  An upgraded Tevatron
with higher luminosity (say $25~fb^{-1}$) may be able to probe  
 a Higgs boson
mass up to about $120~GeV$.\cite{gunsnow,tev2000} This result
depends critically on the $b$ tagging capabilities of the
detectors, since it requires reconstructing the mass of both
$b-$ jets.  

\subsection{Associated Production with Top,
$pp (p {\overline p})\rightarrow t {\overline t} h$.} 

A potentially important mechanism for Higgs production at the LHC
is the associated production with a $t {\overline t}$ pair.
The $t{\overline t}h$ final state can result from either
 $q {\overline q}$
annihilation or from gluon-gluon scattering. 
From the dotted line in
Fig. 8, we see that the rate is roughly $1~pb$ for
$M_h\sim 100~GeV$.  Although the rate is small, requiring an
isolated charged lepton from the top decay 
significantly reduces the background and this decay may be
useful. 
The $t {\overline t}h$ production mechanism is the
only mode for which the QCD radiative corrections have not yet 
been calculated. 
A measurement of this channel will directly probe the $t {\overline t}
h$ Yukawa coupling.  

There are some interesting theoretical problems 
involved in calculating the rate for $t {\overline t}h$
production.\cite{dw}  The dominant contribution is from the
gluon content of the proton.  One can think about the $t {\overline t}$
quarks in the proton fusing to form a Higgs boson or about the
full $gg\rightarrow t {\overline t} h$ process.  Clearly, there is a 
potential for double counting if both the $t {\overline t}$ and
$gg$ fusion production mechanisms are included.  The 
prediction obtained using only the rate for $gg\rightarrow
t {\overline t}h$ will be accurate everywhere except $M_h
>>M_t$, where there will be large logarithms of the form $\log(M_h/M_t)$.
On the other hand, the fusion prediction, $t {\overline t}
\rightarrow h$, will be accurate only if $M_h < M_t$.  The
resolution of the problem is to include the effects of gluon
radiation in the leading logarithm approximation to all
orders in $\alpha_s$ by using the heavy quark distribution functions,
which include this radiation.  However, to ${\cal O}(\alpha_s)$
this logarithm already appears in the $gg\rightarrow t {\overline t}h$
process and hence must be subtracted from the heavy quark distribution
functions.  The answer obtained in this manner is consistent for
all values of $M_h$ and $M_t$ and is shown in Fig. 8 for the LHC.  

The best experimental limit on the Higgs boson mass
at present comes from $Z$ decays at LEP and so we turn now to
a discussion of producing the intermediate mass Higgs
boson in $e^+e^-$ collisions.

\section{Higgs Production in $e^+e^-$ Collisions}
\subsection{Higgs Production in $Z$ decays}    
Since the Higgs boson coupling to the electron is very small, $\sim m_e/v$,
the $s-$channel production mechanism,
 $e^+e^-\rightarrow h$, is miniscule and
the dominant production mechanism
at LEP and LEPII
 for the intermediate mass Higgs boson is
the associated production with a $Z$, 
$e^+e^-\rightarrow Z^*\rightarrow Zh$.

  At the LEP collider with
$\sqrt{s}\sim M_Z$, Higgs production 
could result from the on-shell
decay $Z\rightarrow h f {\overline f} $.  Neglecting
fermion masses and the $Z$ 
boson width, the rate for
Higgs production  from $Z$ decay is given by,\cite{hhg}
\beqn
{BR(Z\rightarrow h f {\overline f})\over
BR(Z\rightarrow f {\overline f})} &=&
{G_FM_Z^2\over 24 \sqrt{2}\pi^2}\biggl\{
{3 y (y^4-8 y^2+20)\over \sqrt{4-y^2}}\cos^{-1}
\biggl({y(3-y^2)\over 2}\biggr)
\nonumber \\
&&-3(y^4-6y^2+4)\log y-{1\over 2}(1-y^2)
(2 y^4-13 y^2+47)\biggr\}
,
\nonumber \\
&&
\eeqn
where $y\equiv M_h/M_Z$.
The branching ratio for $Z\rightarrow h l^+l^-$ is shown in
Fig. 10. It is clear  that
Higgs boson production in $Z$ decays can never be more than
a $1\%$ effect on the total $Z$ width even for a very light
Higgs boson.  

The Higgs boson has been searched
 for in virtual $Z$ decays  at
the LEP collider.
 The primary decay mechanism  used is 
$Z\rightarrow
h l^+l^-$. 
 The decay $Z\rightarrow h \nu {\overline \nu}$ is also useful since the
branching ratio is six times larger than that of
 $Z\rightarrow h l^+l^-$.
The strategy is to search through
 each range of Higgs boson masses separately by
looking for the relevant Higgs decays.
  For example, a light Higgs boson, $M_h< 2 m_e$,
necessarily decays to two photons.  For $M_h\sim 1~MeV$, the Higgs lifetime
is $c \tau\sim 10^3~cm$ and so the Higgs boson is long lived and escapes the
detector without interacting.  
In this case a signal could be
$e^+e^-\rightarrow Z\rightarrow l^+l^- h$ and the signal
is $l^+l^-$ plus missing energy from the undetected Higgs boson.
For each mass region, the appropriate Higgs decay channels are searched for.  
When the Higgs boson becomes heavier than twice the $b$ quark mass, 
it decays primarily
to $b {\overline b}$ pairs and the signal
resulting from leptonic $Z$ decays is then
$e^+e^-\rightarrow Z\rightarrow l^+l^-h\rightarrow l^+l^-$ + jets.  By
a systematic study of Higgs boson masses and decay channels,
the four LEP experiments have found the limit,\cite{leplim} 
\beq
M_h>65.2~GeV \quad .
\eeq
Note that there is no region where light Higgs boson masses are allowed.  The
LEP limits thus obviate early studies of mechanisms such as $K\rightarrow 
\pi h$.
The decay $Z\rightarrow h \gamma$, shown in Fig. 10, has too small
a rate to be useful.  
A review of the LEP limits on the Standard Model Higgs boson is given in
the chapter by Janot.

\begin{figure}[t] 
\vspace*{-1in}  
\centerline{\psfig{figure=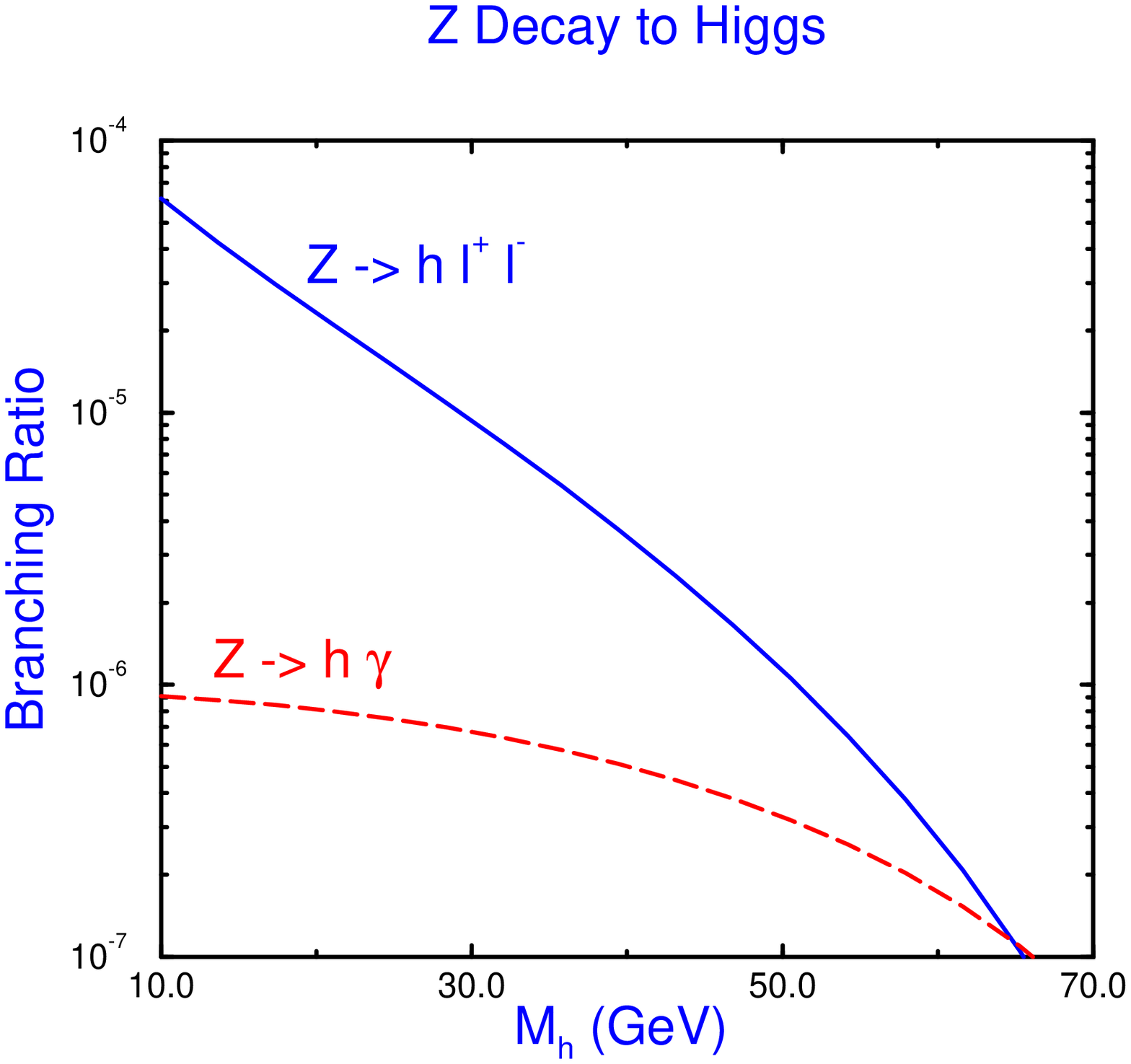,height=3.5in}}
\caption{Branching ratio of the $Z$ boson to the Standard Model
Higgs boson plus a charged lepton pair, $Z\rightarrow
h l^+l^-$,  or  to $Z\rightarrow h \gamma$.}
\label{fig:zhprod}
\end{figure}
\begin{figure}[tb]
\psfig{figure=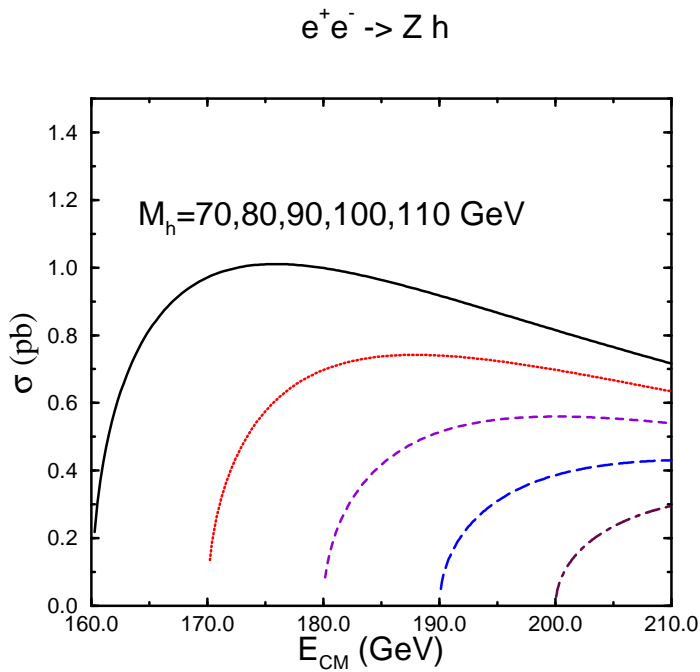,height=3.5in}
\caption{Born cross section for $e^+e^-\rightarrow Zh$ as a
function of the center of mass energy, $E_{CM}$, and for
various values of the Higgs boson mass.}
\end{figure}

\subsection{Associated Production, $e^+e^-\rightarrow Zh$}
At LEPII, the primary production mechanism for the Standard
Model Higgs boson is $e^+e^-\rightarrow Zh$, in which a physical
$Z$  boson is produced.\cite{pet}  The cross section is,
\beq
{\hat \sigma}(e^+e^-\rightarrow h Z)={\pi \alpha^2\lambda_{Zh}^{1/2}
[\lambda_{Zh}+12 {M_Z^2\over s}](1+(1-4 \sin^2\theta_W)^2]
\over 192 s \sin^4\theta_W\cos^4\theta_W(1-M_Z^2/s)^2}
\label{eezhborn}
\eeq
where
\beq
\lambda_{Zh}\equiv (1-{M_h^2+M_Z^2\over s})^2-{4 M_h^2M_Z^2
\over s}
\quad .
\eeq
The center of mass momentum of the produced $Z$ is $
\lambda_{Zh}^{1/2}\sqrt{s}/2$ and  the cross section is
shown in Fig. 11  as a function of $\sqrt{s}$ for  different
values of the Higgs boson mass.  The cross section
peaks at $\sqrt{s}\sim M_Z+ 2 M_h$.  
  
The electroweak radiative
corrections are quite small at LEPII energies.\cite{kniehl}
Photon bremsstrahlung can be important however since it
is enhanced by a large logarithm, $\log(s/m_e^2)$.  The photon radiation
can be accounted for by integrating
the Born cross section of Eq. \ref{eezhborn}  with a radiator function
$F$
which includes virtual and soft photon effects, along with
hard photon radiation,\cite{berends} 
\beq
\sigma={1\over s} \int  ds^\prime F(x,s)
{\hat \sigma}(s^\prime)
\eeq
where $x=1-s^\prime/s$ and
the radiator function $F(x,s)$ is known to ${\cal O}(\alpha^2)$, along
with the exponentiation of the infrared contribution,  
\beqn 
F(x,s)&=& tx^{t-1}\biggl\{1+{3\over 4} t\biggr\}
+\biggl\{{x\over 2}-1\biggr\} t+{\cal O}(t^2) 
\nonumber \\
t&\equiv & {2\alpha\over
\pi}\biggl[ \log\biggl({s\over m_e^2}\biggr) -1\biggr]
\quad .
\eeqn  
Photon radiation significantly reduces
  the $Zh$ production
rate from the Born cross section
as shown in Fig. 12. 
\begin{figure}[tb]
\psfig{figure=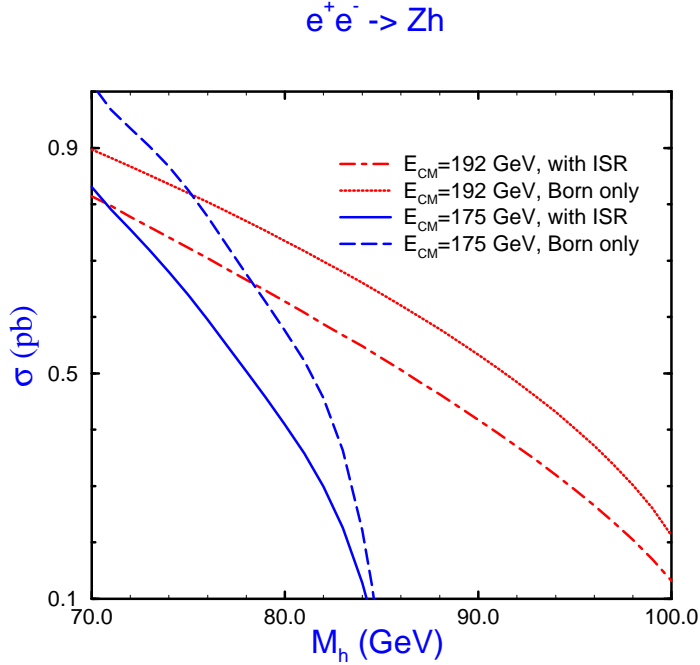,height=3.5in}
\caption{Effects of initial state radiation(ISR) on the process $e^+e^-
\rightarrow Zh$. The curves labelled ``Born only'' are the
results of Eq. 52, while those labelled  ``with ISR'' include
the photon radiation as in Eq. 54.}
\end{figure}

The
intermediate mass
 Higgs boson will decay mostly to $b {\overline b}$ pairs,
so the final state from $e^+e^-\rightarrow Zh$
 will have four fermions.
The dominant background
 is $Z b {\overline b}$ production, which can
be efficiently eliminated by $b$-tagging
 almost up 
to the kinematic limit for producing the Higgs boson.  
LEPII studies estimate that with $\sqrt{s}=184~GeV$ and 
${\cal L}=150 pb^{-1}$, a Higgs boson mass of $87~GeV$
could be observed at the $5\sigma$ level.\cite{lepstud}
  With higher energy, $\sqrt{s}=192~GeV$
and the same luminosity, masses up to $95~GeV$ could be
reached.   
A higher energy $e^+e^-$ machine
(such as an NLC with $\sqrt{s}\sim 500~GeV$)
 could push the Higgs mass limit
to around $M_h\sim .7 \sqrt{s}$.  

Currently LEPII has data from both $\sqrt{s}=161~$ and 
$\sqrt{s}=172~GeV$.  
ALEPH has announced a preliminary limit on the Higgs boson
mass of~\cite{lep2lim}
\beq
M_h> 70.7~GeV
\quad 
\eeq
from $10 pb^{-1}$ of data.
This analysis includes both hadronic and leptonic decay modes of
the $Z$.  Note how close this limit is
to the kinematic boundary.

          It is interesting to compare the upper bound on
the Higgs mass from LEPII with the
lowest mass reach of the $h\rightarrow \gamma\gamma$ process
at the LHC.  At LEPII, with $\sqrt{s}=192~GeV$,
a mass of $M_H\sim 95~GeV$ will be probed,
while the LHC will observe down to $M_h\sim 85~GeV$
in the $\gamma\gamma$ decay mode.
We expect therefore that there will be no mass gap in the
Higgs mass coverage, with the results from LEPII neatly meshing
with those from the LHC. The higher energy at LEPII,
 $\sqrt{s}\sim 192~GeV$,
is obviously necessary for this to be the case.

The cross section for $e^+e^-\rightarrow Zh$ is $s$-wave and
so has a very steep dependence on energy and on the Higgs boson
mass at threshold, as can be seen  clearly in  Fig. 11.
This  makes possible a precision measurement of the Higgs
mass.
 Reconstructing the
final state momenta at
 an NLC with $\sqrt{s}=500~GeV$ and assuming
an SLD like detector could give
 a mass measurement with an accuracy of
\cite{dem2}
\beq
\Delta M_h\sim 180~MeV \sqrt{{50~fb^{-1}\over L}}
\quad .
\eeq
By measuring the cross section at threshold and
normalizing to a second measurement above threshold in order
to minimize systematic uncertainties, a $1\sigma$
measurement of the mass can be obtained~\cite{bbgh}
\beq
\Delta M_h\sim 60\sqrt{{100 fb^{-1}\over L}}
\quad\quad {\rm for~~}M_h=100~GeV       \quad ,
\eeq
where $L$ is the total integrated luminosity.  
The precision becomes worse for larger
 $M_h$ because of the
decrease in the signal cross section.  (Note that the luminosity
at LEPII will not be high enough to perform this measurement.) 

 The angular distribution of the Higgs boson from the
$e^+e^-\rightarrow Zh$ process  is
\beq
{1\over\sigma}
{d\sigma\over d\cos\theta}\sim \lambda_{Zh}^2
\sin^2\theta+{8M_Z^2\over s}
\eeq
so that at high energy the distribution is
that of a scalar particle,
\beq
{1\over\sigma}{d\sigma\over d \cos\theta}\rightarrow {3\over 4}\sin^2
\theta
\quad .
\eeq  
If the Higgs boson were CP odd, on the other hand, the angular
distribution would be $1+\cos^2\theta$. Hence the angular distribution
is sensitive to the spin-parity assignments of the Higgs boson.

\subsection{Higgs Production in Vector Boson Fusion, $VV\rightarrow h$}
 
In $e^+e^-$ collisions
the Higgs boson can also be produced by $W^+W^-$ fusion,\cite{hhg,dr}
\beq
e^+e^-\rightarrow W^+W^- \nu {\overline \nu}
\rightarrow h \nu {\overline \nu},
\eeq
 and by
$ZZ$ fusion,
\beq
e^+e^-
\rightarrow ZZe^+e^-\rightarrow he^+e^-
.
\eeq
The fusion cross sections are easily found,~\cite{abdel} 
\beq
\sigma_{VVh}={G_F^3M_V^4\over 64\sqrt{2}\pi^3}
\int^1_{{M_h^2\over s} } dx \int^1_x {dy\over
(1+s(y-x)/M_V^2)^2}
\biggl((v^2+a^2)^2f(x,y)+4v^2a^2g(x,y)\biggr)
\eeq
where,
\beqn  
f(x,y)&=&\biggl({2x\over y^3}-{1+2x\over y^2}
+{2+x\over 2 y}-{1\over 2}\biggr)\biggl(
{w\over 1+w}-\log(1+w)\biggr)+{x\over y^2}{w^2(1-y)\over 1+w}
\nonumber \\
g(x,y)&=& \biggl( -{x\over y^2}+{2+x\over 2 y}-{1\over 2}
\biggr)\biggl({w\over 1+w}-\log(1+w)\biggr)
\nonumber \\
w&\equiv & {y(sx-M_h^2)\over M_V^2  x}
\eeqn 
 and $v=a=\sqrt{2}$ for $e^+e^-\rightarrow W^+W^-\nu
\overline{\nu}\rightarrow \nu{\overline\nu} h$  
 and $v=-1+4 \sin^2\theta_W,~a=-1$
for $e^+e^-\rightarrow ZZe^+e^-\rightarrow e^+e^-h$.
The vector boson fusion cross sections are shown in Fig. 13 as
a function of $\sqrt{s}$.  
\begin{figure}[tb]
\centerline{\epsfig{file=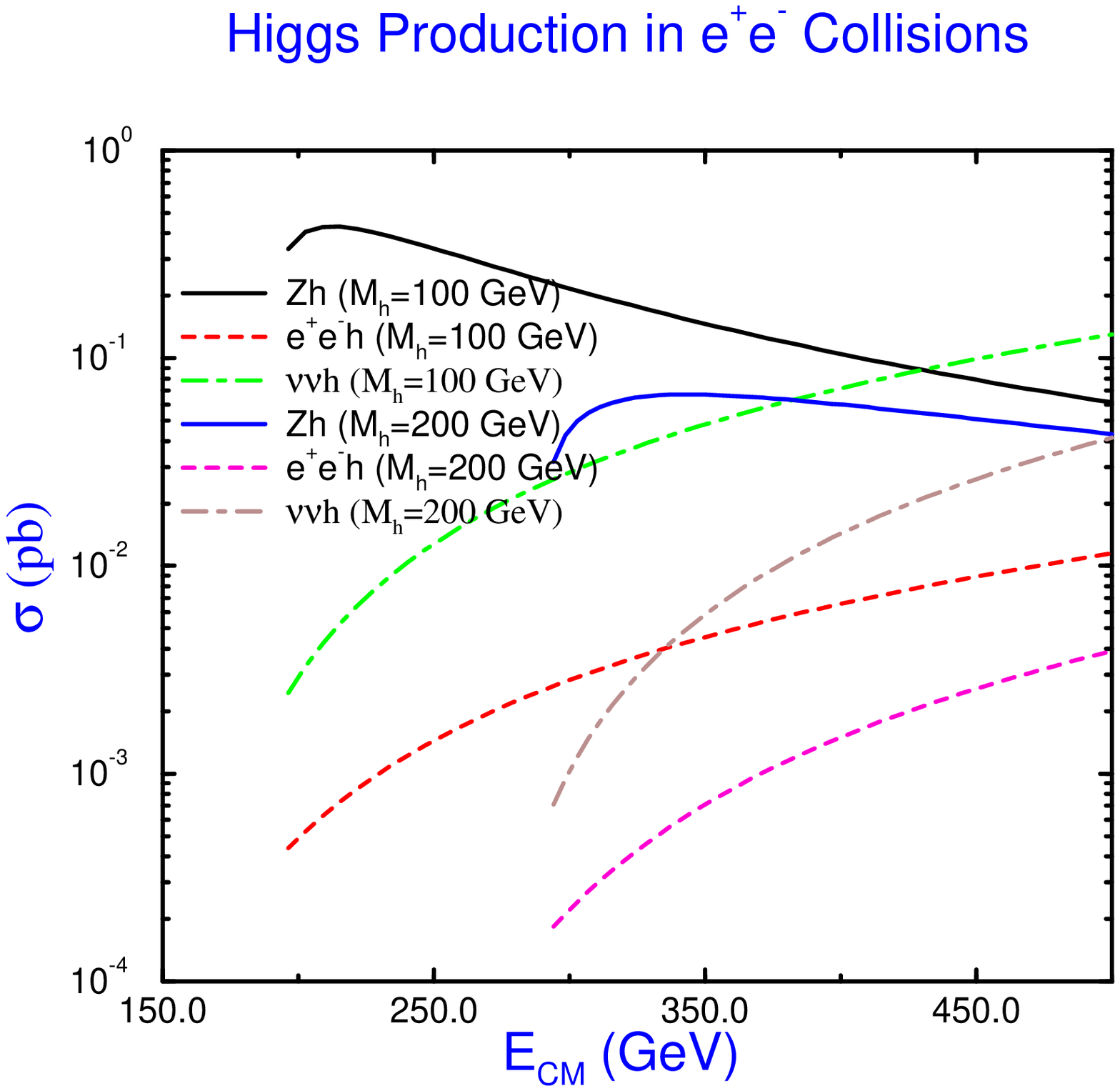,height=3.5in}}
\caption{Higgs boson production in $e^+e^-$ collisions
for $M_h=100$ and $200~GeV$.}
\end{figure} 
The $ZZ$ fusion cross section is an order of magnitude smaller than the
$W^+W^-$ fusion process due to the smaller neutral current couplings.
This suppression
is partially compensated for experimentally
 by the fact that the $e^+e^-h$ final
state permits a missing mass analysis to determine the Higgs mass.

At an NLC, the cross section for vector boson fusion,
 $e^+e^-\rightarrow W^+W^-\nu {\overline \nu}
\rightarrow h \nu {\overline \nu}$ and $e^+ e^-\rightarrow Zh$ are
of similar  size for a $100~GeV$ Higgs boson.   
The fusion processes grow as $(1/M_W^2)\log(s/M_h^2)$, while
the $s-$ channel process, $e^+e^-\rightarrow Zh$, falls as
$1/s$ and so at high enough energy the fusion process will
dominate, as can be seen in Fig. 13. 

\subsection{$e^+e^-\rightarrow t {\overline t}h$}  
Higgs production in association with a $t {\overline t}$ pair is small at
an $e^+e^-$ collider.  At $\sqrt{s}=500~GeV$, $20~fb^{-1}$ of
luminosity would produce only $20$ events for $M_h=100~GeV$.  
The signature for this
final state would be spectacular, however, since it would 
predominantly be $W^+W^- b {\overline b} b {\overline b}$, which
would have a very small background.
 The $t {\overline t}h$ final state results almost 
completely from Higgs bremsstrahlung off the top quarks and could 
potentially yield a direct measurement of the $t {\overline t}h$
Yukawa coupling.\cite{kal}  
 
\section{Higgs Production in $\mu^+\mu-$ Collisions}
An intermediate mass Higgs boson can also be probed
via $\mu^+\mu^-\rightarrow Zh$ and the physics of this
mechanism is almost identical to that of $e^+e^-\rightarrow
Zh$.  However, 
 a $\mu^+\mu^-$ collider with high luminosity and narrow beam
spread also
offers the possibility of performing high precision measurements of
the Higgs boson mass.
Since the Higgs boson coupling is proportional to mass,
the $s-$ channel process $\mu^+\mu^-\rightarrow h$ is considerably larger
than the corresponding process in an $e^+e^-$ collider.
The cross section for the resonant process $\mu^+\mu^-\rightarrow
h\rightarrow X$ is,  
\beq 
\sigma_h(s)={4 \pi \Gamma(h\rightarrow \mu^+\mu^-)\Gamma(h
\rightarrow X)\over
(s-M_h^2)^2+M_h^2\Gamma_{tot}^2}
\quad ,
\label{mures}
\eeq
where $\Gamma_{tot}$ is the total decay width of the Higgs boson.
This process is obviously maximized when $\sqrt{s}\sim M_h$.  One
envisions discovering the Higgs boson via
$e^+e^-\rightarrow Zh$ (or $\mu^+\mu^-\rightarrow Zh$) and
then building a storage ring for a muon collider such that
$\sqrt{s}\sim M_h$.
At such a machine, precision studies of the Higgs mass and couplings
could be performed.~\cite{bbgh} 

The crucial question is whether the energy spread of the beam is greater
or smaller than the intrinsic width of the Higgs boson.  Because the
muon is so much heavier than the electron, there will be less initial
state radiation and the beam energy resolution will typically be 
better in a muon collider than in
an $e^+e^-$ collider.  The beam energy resolution can be 
parameterized as a Gaussian with an rms deviation, $R$.
This leads to an energy resolution, $\delta E$, of,
\beq
\delta E\sim 30~MeV\biggl({R\over .05\%}\biggr)\biggl({\sqrt{s}\over 100}
\biggr) .
\label{den} 
\eeq
For a $\mu^+\mu^-$ collider, one expects values of
$R\sim .1$ to $.01\%$, while for an $e^+e^-$ collider,
$R\sim 1\%$. 
 For $M_h<150~MeV$, the energy spread of the beam will be
smaller than the Higgs width if $R<.05\%$, (See Fig. 4).

The physical cross section is then
found by convoluting Eq. \ref{mures}
with the energy resolution,  
\beq
\sigma(s)=\int \sigma_h(s^\prime) d\sqrt{s^\prime} exp^{
-(\sqrt{s^\prime}-\sqrt{s})^2/(2 \delta E)
}  
{1\over  \sqrt{2\pi} \delta E}
\quad .
\eeq 
For $\delta E <<\Gamma_{tot}$, the cross section is 
\beq
\sigma(s)\sim {\sqrt{2\pi}\pi\over M_h^2 \delta E}
\Gamma(h\rightarrow\mu^+\mu^-)
{\Gamma(h\rightarrow X)\over \Gamma_{tot}}
\quad .
\eeq 
The $s$-channel Higgs cross section is shown in Fig. 14 
for various values of $R$.  
The increase in the cross section with decreasing $R$ is clearly seen.  

Detailed studies have estimated that with $R\sim .01\%$
and $7$ scan points centered around $\sqrt{s}\sim M_h$
(with a total luminosity of $3.5~fb^{-1}$)
 a measurement~\cite{bbgh}
\beq
\Delta M_h\sim 4~MeV
\quad  
\eeq 
can be obtained.  
This is an order of magnitude smaller than that obtainable at
an NLC (see Eq. 59).
The height of the peak in Fig. 14 is a measure of $\Gamma_{tot}$ and 
the same $7$ scan points would yield a $10\%$ measurement
of $\Gamma_{tot}$.
A measuremtns of $\Gamma_{tot}$ is an important discriminator
between the Standard Model Higgs boson and other non-Standard
 Model scalars.

\begin{figure}[tb]
\epsfig{file=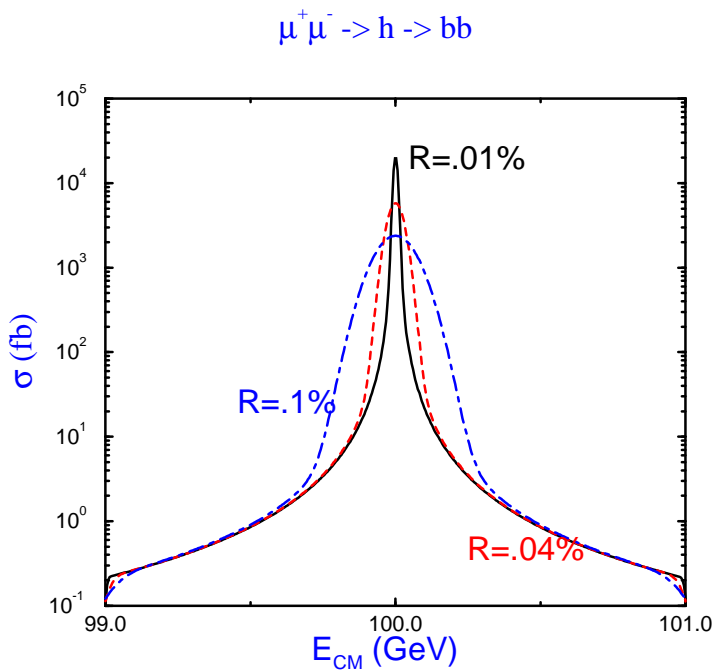,height=3.5in}
\caption{Cross section for resonant Higgs production in
a $\mu^+\mu^-$ collider with $M_h=100~GeV$,
 for various values of the rms
deviation in the energy spread, $R$.}
\end{figure}
\section{Higgs Production in $\gamma\gamma$ Collisions}  
 
It is possible that an NLC with
$\sqrt{s}\sim 500~GeV$ will be able to use back-scattered
lasers to produce $\gamma\gamma$ 
or $e\gamma$ collisions with high energy
and high luminosity.  The $\gamma \gamma$
collisions may be useful for discovering the intermediate
mass Higgs boson since  the Higgs production
rate is proportional to $\Gamma(h\rightarrow \gamma\gamma)$
and the full $\gamma\gamma$ center of mass energy goes into
creating the Higgs boson.

The process we imagine observing is $\gamma \gamma\rightarrow
h \rightarrow b {\overline b}$.  The dominant background
from $\gamma\gamma\rightarrow b {\overline b}$ is relatively
easy to control.  The signal is produced in a $J_Z=0$ state, while the
background is mostly from $J_Z=2$.  Therefore, polarizing the photons will 
efficiently discriminate against the background.  There
is also a significant background from ``resolved photons'', where
the effective process is $\gamma g$ or $gg\rightarrow b {\overline b}$.
These resolved backgrounds are more severe for lower invariant
masses and so we expect the lighter Higgs masses to be the
hardest to see.  Ref. \cite{gamgam} estimates that
$10 fb^{-1}$ will allow a Higgs signal to
be extracted in the range $110~ GeV < M_h < 140~ GeV$
at a $500~GeV$ NLC.

A $\gamma\gamma$ collider will allow a direct measurement
of the $h\gamma\gamma$ coupling, which is sensitive to all 
charged particles coupling to the Higgs boson.  This coupling
is also sensitive to the $W^+W^-\gamma$ and
$W^+W^-h$ couplings and so provides a window to non-Standard
Model gauge interactions.

An intermediate mass Higgs boson may also be probed in heavy
ion collisions through the coherent $2$ photon production in
the electromagnetic field of a nucleus with charge $Z$.  Such
interactions are enhanced by a factor of $Z^4$.  Using
calcium beams at the LHC, it may be possible to obtain
a $3-4 \sigma$ signal for a Higgs boson in the range
$100 < M_h < 130~GeV$.\cite{papa} 
\section{Conclusion}

The theoretical motivations for a Higgs boson in the intermediate
mass region are  extremely compelling, making it vital that
this region be probed experimentally.
If the Standard Model is the theory of electroweak interactions
at energies 
all the way to the Planck scale, then the Higgs boson must
exist in this region.  Indirect electroweak measurements also
support the validity of the intermediate mass hypothesis.  
  
 The combination of LEPII and the LHC should suffice to
establish the existence of the Higgs boson if it is in
the intermediate mass regime, although the region between $80$ and $90$ 
GeV is extremely challenging experimentally.  
The Higgs boson, once discovered,
 must also be measured in a variety of production
channels and decay modes in order to confirm the Standard Model couplings.

\section*{Acknowledgments}

This manuscript has been authored under contract number DE-AC02-76CH00016
with the U.S. Department of Energy.  Accordingly, the U.S. Government retains a non-exclusive, royalty free license to publish or
reproduce the published form of this contribution, or allow others to do so,
for U.S. government purposes.


\section*{References}
\begin{thebibliography}{99}
\bibitem{leplim}J. Grivas,{\it  European Physical Society International
Europhysics Conference on High Energy Physics}, Brussels, Belgium,
July 27 - August  2, 1995.

\bibitem{lep2lim} G. Cowen, CERN seminar, Feb. 25, 1997,
http: //alephwww.cern.ch /ALPUB /seminar /Cowan-172-jam /cowan.html.

\bibitem{lat}  A.~Hasenfratz, {\it Quantum Fields on the
Computer},  (World Scientific, Singapore, 1992),
ed. M.~Creutz.

\bibitem{unit}  B.Lee, C. Quigg, and H. Thacker, {\it Phys.
Rev.} {\bf D16} (1977) 1519;  D. Dicus and V. Mathur, {\it
Phys. Rev.} {\bf D7} (1973) 3111.

\bibitem{susymass}   P. Chankowski, S. Pokorski, and
J. Rosiek, {\it Phys. Lett.} {\bf B274} (1992) 191;
J. Espinosa and M. Quiros, {\it Phys. Lett.}
{\bf BB267} (1991) 27;
H.~Haber and R. Hempfling, {\it Phys. Rev.} {\bf D48}
(1993) 4280; J.~Ellis, G.~Ridolfi, and F.~Zwirner, 
{\it Phys. Lett.} {\bf B257} (1991) 83.
  See also the chapter by 
P. Chankowski and S. Pokorski, hep-ph/9702431. 

\bibitem{kkw} C. Kolda, G.~Kane, and J.~Wells, {\it Phys.
Rev. Lett.} {\bf 70} (1993) 2686.

\bibitem{triv} N.Cabibbo, L. Maiani, G. Parisi, and R. Petronzio,
{\it Nucl. Phys.} {\bf B158} (1979) 295; M. Sher, {\it Phys. Rev.}
 {\bf 179} (1989) 273;
L.~Maiani, G.~Parisi, and R.~Petronzio, {\it Nucl. Phys.}
{\bf B136} (1978) 115;
M.Lindner, {\it Z. Phys.} {\bf C31} (1986) 295;
M.~Lindner, M.~Sher, and H.~Zaglauer, {\it Phys. Lett.}
{\bf 228} (1989) 139.


\bibitem{sher}M.~Sher, {\it Phys. Lett.} {\bf B317} (1993) 159;
{\bf B331} (1994) 448; G. Altarelli and I. Isidori,
{\it Phys. Lett.} {\bf B337} (1994) 141;
J. Casa, J. Espinosa, and M. Quiros, {\it Phys. Lett.}
{\bf B342} (1995) 171; C.~Ford, D.~Jones, P.~Stephenson, and
M. Einhorn, {\it Nucl. Phys.} {\bf B395} (1993) 17.  


\bibitem{eqpap} J.~Espinosa and M.~Quiros, {\it Phys. Lett.}
{\bf 353B} (1995) 257.
\bibitem{dem2} U. Baur, M. Demarteau {\it et. al.}, {\it
Proceedings of the 1996 Snowmass Workshop}, Snowmass,
CO, June, 1996, hep-ph/9611334.                    


\bibitem{rosner} J. Rosner, Lectures given at
{\it The Cargese Summer Institute on Particle Physics}, 1996,
hep-ph/9610222.   


\bibitem{ellis} J.~Ellis, G.~Fogli, and E. Lisi, {\it
Phys. Lett.} {\bf B333} (1994) 118.
 
\bibitem{gsw} J.~Gunion, A. Stange, and S.~Willenbrock, 
to be published in {\it Electroweak Symmetry
Breaking and Beyond the Standard Model}, (World Scientific,
Singapore, 1997), ed. T. Barklow {\it et. al.}, hep-ph/9602238.

\bibitem{gunsnow} J.~Gunion {\it et. al.},
 {\it
Proceedings of the 1996 Snowmass Workshop}, Snowmass,
CO, June, 1996, hep-ph/9703330.                    

\bibitem{kniehl} B.~Kniehl, {\it Phys. Rept.} {\bf 240} (1994) 211.

\bibitem{abdel} A. Djouadi, {\it Int. Journ. Mod. Phys.} 
{\bf A10} (1995) 1, hep-ph/9406430.  


\bibitem{spirev} M.~Spira, A.~Djouadi, D. Graudenz, and P.~Zerwas,
{\it Nucl. Phys.} {\bf B453} (1995) 17.

\bibitem{hbqcd} E. Braaten and J. Leveille,
{\it Phys. Rev.} {\bf D22} (1980) 715;
N.~Sakai, {\it Phys. Rev.} {\bf D22} (1980) 2220;
T. Inami and T. Kubota, {\it Nucl. Phys. } {\bf B179}
(1981) 171;
M. Drees and K.Hikasa, {\it Phys. Lett.} {\bf B240}
(1990) 455;
E. Gross, B. Kniehl and G. Wolf, {\it Z. Phys.} {\bf C63}
(1994) 417;
{\bf C66} (1995) 321 {\bf E}.

\bibitem{dsz}
A. Djouadi, M. Spira, and P. Zerwas, {\it Z. Phys.} {\bf C70}
(1996) 427; {\it Phys. Lett} {\bf B264} (1991) 440.  


   
\bibitem{h2l}{S.Gorishny, A. Kataev, S. Larin, and L. Surguladze,
{\it Mod. Phys. Lett.} {\bf A5} (1990) 2703.}

\bibitem{ewh} J.~Fleischer and F. Jegerlehner, {\it Phys. 
Rev.} {\bf D23} (1981) 2001;
A. Dabelstein and W. Hollik, {\it Nucl. Phys.} {\bf C53} (1992) 507;
B. Kniehl, {\it Nucl. Phys.} {\bf B376} (1992) 3. 

\bibitem{hdecay} M.Spira, {\it Proceedings~ of~ AIHENP~ 96},
 Lausanne,
Switzerland, Sept. 1996, CERN-TH-96-285, hep-ph/9610350.


\bibitem{keith} R. Ellis {\it et. al.}, {\it Nucl. Phys.} {\bf B297}
(1988) 221.

\bibitem{hhg} J. Gunion {\it et.al.}, {\it The Higgs Hunter's
Guide} (Addison Wesley, Menlo Park, 1990.) 

\bibitem{hggg}T. Inami, T. Kubota, and Y. Okada, {\it Z. Phys.}
{\bf C18} (1983) 69.


\bibitem{wkwm} W.-Y.~Keung and W.~Marciano, {\it Phys.
Rev.} {\bf D30} (1984) 248.  

\bibitem{hzg}  L.~Bergstrom and G.~Hulth, {\it Nucl. Phys.}
{\bf B259} (1985) 137;
 R. Cahn, M. Chanowitz, and N. Fleishon,
{\it Phys. Rev.} {\bf B82} (1979) 113.  

\bibitem{hgg} A.~Vainshtein, M.~Voloshin, V.~Zakharov, and
	M.~Shifman, {\it Sov. J. Nucl. Phys.} {\bf 30} (1979) 711;
	M.~Okun, {\it Leptons and Quarks}, (North-Holland,
	Amsterdam, 1982).
 
  
  
\bibitem{spgg} A.~Djouadi, M.~Spira, and P.~Zerwas, {\it Phys.
	Lett.} {\bf B311} (1993) 255; S.~Dawson and R. Kauffman,
	{\it Phys. Rev.} {\bf D47} (1993) 1264.  

\bibitem{hqcd} D. Graudenz, M. Spira and P. Zerwas, {\it
Phys. Rev. Lett.}{\bf 70} (1993) 1372. 



\bibitem{daw}
S. Dawson, {\it Nucl. Phys.} {\bf B359} (1991) 283.

\bibitem{dg} A.~Djouadi and P. Gambino, {\it Phys. Rev.
Lett.} {\bf 73} (1994) 2528

\bibitem{qcdsum} R . Kauffman, {\it Phys. Rev. } {\bf D44} (1991) 1415;
 I. Hinchliffe and S. Novaes, {\it Phys. Rev. } {\bf D38} (1988) 3475;
	C. Yuan, {\it Phys. Lett.} {\bf B283} (1992) 395.  

  
\bibitem{colsop} J.~Collins and D.~Soper, {\it Nucl. Phys.}
{\bf B193} (1981) 381; {\bf B213} (1983) 545 {\bf E};
{\bf B197} (1982) 446;
J.~Collins, D. Soper, and G.~Sterman, {
\it
Nucl. Phys. } {\bf B250} (1985) 199.  

\bibitem{soft} M. Kramer, E. Laenen, and M. Spira,
	CERN-TH-96-231, hep-ph/9611272.

\bibitem{stirl} Z. Kunszt, S.~Moretti, and W.~Stirling,
DFTT-34/95, hep-ph/9611397.


\bibitem{atcms} ATLAS Technical Proposal, CERN/LHCC/94-43,
	LHCC/P2 (1994);  CMS Technical Proposal, CERN/LHCC/94-38,
	LHCC/P1 (1994).  

\bibitem{gwud} J.~Gunion, G.~Kane, and J.~Wudka, {\it Nucl. Phys.}
{\bf B299} (1988) 23.


\bibitem{hw} T. Han and S. Willenbrock, {\it Phys. Lett.}
{\bf B273} (1991) 167.

\bibitem{dy} G. Altarelli, R. Ellis, and G. Martinelli, {\it Nucl
Phys. } {\bf B157} (1979) 461; J. Kubar-Andre and F. Paige,
{\it Phys. Rev.} {\bf D19} (1979) 221. 
  
\bibitem{stange} A.~Stange, W. Marciano, and S.~Willenbrock,
{\it Phys. Rev.} {\bf D50} (1994) 4491; {\bf D49} (1994) 1354.

\bibitem{tev2000}  {\it Report of the Tev2000 Study Group on Future
Electroweak Physics at the Tevatron}, eds. D. Amidei and R. Brock;
FERMILAB-PUB-96-082, 1996. 

\bibitem{mren} S.~Mrenna and C.~Yuan, hep-ph/9703224.

\bibitem{dw} D.~Dicus and S.~Willenbrock, {\it Phys. Rev.}
{\bf D39} (1989)751; Z.~Kunszt, {\it Nucl. Phys.}
{\bf 
B247} (1984) 339.


\bibitem{pet} D. Jones
 and S. Petcov, {\it Phys. Lett.} {\bf B84} (1979) 440.
  


\bibitem{berends} F. Berends, {\it $Z$ physics at LEP1}, CERN Yellow
Report No 89-08, Geneva, 1989, Vol. 1, ed., G. Altarelli,
R. Kleiss, and C. Verzegnassi; F. Berends, G. Burgess, and
W. van Neervan, {\it Nucl. Phys.} {\bf B297} (1988) 429;
{\bf B304} (1988) 921 {\bf E} .



\bibitem{lepstud} M. Carena, P. Zerwas, {\it et. al.},
{\it Higgs Physics}, hep-ph/9602250.

\bibitem{bbgh} V. Barger, M. Berger, J. Gunion, and T. Han,
hep-ph/9612279; hep-ph/9602416;
{\it Phys. Rev. Lett.} {\bf75} (1995) 1462. 

\bibitem{dr} S. Dawson and J. Rosner, {\it Phys. Lett.}
{\bf B148} (1984) 497.


\bibitem{kal} A.~ Djouadi, J.~Kalinowski and P.~Zerwas,
{\it Z. Phys.} {\bf C54} (1992) 255.  


\bibitem{gamgam} J.~Gunion and H. ~Haber, {\it Phy. Rev.}
{\bf D48} (1993) 5109;  
M. Baillargeon, G. Belanger, and F. Boudjema, Proceedings of
{\it Two Photon Physics from DA$\Phi$NE to LEP200 and Beyond},
Feb. 2-4, 1994, Paris, hep-ph/9405359.    

\bibitem{papa} E.~Papageorgiu, Proceedings of {\it 30th Rencontres do
Moriond}, Meribel les Allues, France, 1995, hep-ph/9507221.

\end{thebibliography}
\end{document}